\documentclass[10pt,journal,compsoc]{IEEEtran}
\ifCLASSOPTIONcompsoc %chktex 1
  \usepackage[nocompress]{cite}
\else
  \usepackage{cite}
\fi
\ifCLASSINFOpdf %chktex 1
  \usepackage[pdftex]{graphicx}
  \graphicspath{{images/}}
  \DeclareGraphicsExtensions{.pdf,.jpeg,.png,.svg,.eps}
\else
\fi
\usepackage{amsmath}
\usepackage{amsthm}
\usepackage[capitalize]{cleveref}
\crefname{algocf}{Alg.}{Algs.}
\Crefname{algocf}{Algorithm}{Algorithms}

\usepackage{mathtools}
\DeclarePairedDelimiter\ceil{\lceil}{\rceil}

\newcommand{\idx}[1]{%
\ifcase#1\relax\ensuremath{i}% Case 0.
\or{}\ensuremath{j}% Case 1.
\or{}\ensuremath{h}% Case 2.
\or{}\ensuremath{n}% Case 3.
\or{}\ensuremath{z}% Case 4.
\or{}\ensuremath{m}% Case 5.
\else\ensuremath{\idx{0}}
\fi
}

\usepackage[linesnumbered, ruled]{algorithm2e}
\SetKwRepeat{Do}{do}{while}
\SetKwInOut{Input}{Input}
\SetKwInOut{Output}{Output}
\SetKwProg{try}{try}{:}{}
\SetKwInOut{KwInit}{Initialization}
\SetKwProg{catch}{catch}{:}{end}

\usepackage{booktabs}
\usepackage[style=index,nonumberlist,nopostdot]{glossaries}

\theoremstyle{definition}
\newtheorem{definition}{Definition}
\newtheorem{corollary}{Corollary}[definition]

\newtheorem{assumption}{Assumption}
\newtheorem{constraint}{Constraint}

\usepackage{multicol}

\glssetwidest{\glsfindwidesttoplevelname{}}
\glsfindwidesttoplevelname{}
\makeglossaries{}
\newglossaryentry{deterministic-safe-set}
{
    name=Deterministic Safe Set,
    description={
            For a given \gls{unique-nodes-found-in-gathering} containing \gls{desired-malicious-node-tolerance} malicious nodes, where \(\gls{amount-of-seen-nodes} > \gls{desired-malicious-node-tolerance}\), \gls{dss} is a deterministic honest set which contains at least one honest node and its size is at least \(\gls{desired-malicious-node-tolerance}+1\)
        }
}

\newglossaryentry{deterministic-progress-set}
{
    name=Deterministic Progress Set,
    description={
            For a given \gls{unique-nodes-found-in-gathering} containing \gls{desired-malicious-node-tolerance} malicious nodes, where \(\gls{amount-of-seen-nodes} > 2\gls{desired-malicious-node-tolerance}\), \gls{dps} is a deterministic honest set which contains a majority of honest nodes and its size is at least \(2\gls{desired-malicious-node-tolerance}+1\)
        }
}

\newglossaryentry{dss}
{
    name=\ensuremath{\Delta_{s}},
    description={
            Deterministic Safe Set
        }
}

\newglossaryentry{dps}
{
    name=\ensuremath{\Delta_{p}},
    description={
            Deterministic Progress Set
        }
}

\newglossaryentry{dsh}
{
name=\ensuremath{\gls{ds}_{\idx{2}}},
description={
        Deterministic Honest Set
    }
}

\newglossaryentry{ds}
{
    name=\ensuremath{\Delta},
    description={
            A shorthand for \gls{dsh} when \idx{2} is implied from context or not relevant
        }
}
\newglossaryentry{ps}
{
    name=\ensuremath{\Pi},
    description={
            A shorthand for \gls{psh} when \idx{2} is implied from context or not relevant
        }
}

\newglossaryentry{psh}
{
name=\ensuremath{\gls{ps}_{\idx{2}}},
description={
        Probabilistic Honest Set
    }
}

\newglossaryentry{probabilistic-safe-set}
{
    name=Probabilistic Safe Set,
    description={
            For a given \gls{unique-nodes-found-in-gathering} containing \gls{desired-malicious-node-tolerance} malicious nodes, where \(\gls{amount-of-seen-nodes} > \gls{desired-malicious-node-tolerance}\), \gls{pss} is a probabilistic honest set which contains at least one honest node with probability \gls{correctness-probability}
        }
}

\newglossaryentry{probabilistic-progress-set}
{
    name=Probabilistic Progress Set,
    description={
            For a given \gls{unique-nodes-found-in-gathering} containing \gls{desired-malicious-node-tolerance} malicious nodes, where \(\gls{amount-of-seen-nodes} > 2\gls{desired-malicious-node-tolerance}\), \gls{pps} is a probabilistic honest set which contains a majority of honest nodes with probability \gls{correctness-probability}
        }
}

\newglossaryentry{correctness-probability}
{
    name=\ensuremath{\rho},
    description={
            Probability guarantee on probabilistic sets correctness
        }
}
\newglossaryentry{first-contact}
{
    name=\ensuremath{fc},
    description={
            First contact node identifier
        }
}

\newglossaryentry{desired-malicious-node-tolerance}
{
    name=\ensuremath{\kappa},
    description={
            Desired malicious node tolerance, expressed as an absolute number
        }
}
\newglossaryentry{transaction-id}
{
    name=\ensuremath{tx_{id}},
    description={
            A transaction identifier
        }
}
\newglossaryentry{block-id}
{
    name=\ensuremath{blk_{id}},
    description={
            A block identifier
        }
}

\newglossaryentry{merkle-root}
{
    name=\ensuremath{m_{root}},
    description={
            A Merkle root
        }
}

\newglossaryentry{chosen-bootstrap-node}
{
    name=\ensuremath{b_{c}},
    description={
            Chosen bootstrap candidate
        }
}
\newglossaryentry{state-of-transaction-inclusion}
{
    name=\ensuremath{tx_{incl}},
    description={
            State of transaction inclusion, can be \textbf{true}, \textbf{false} or ambiguous when the query is unsafe
        }
}

\newglossaryentry{amount-of-seen-nodes}
{
    name=\ensuremath{|\gls{unique-nodes-found-in-gathering}|},
    description={
            The number of unique nodes discovered during a \gls{gathering}, also the population size used for hypergeometric set construction
        }
}

\newglossaryentry{honest-node-amount-in-probabilistic-set}
{
    name=\ensuremath{numHon},
    description={
            Constraint on honest node amount in a probabilistic set \([\lfloor \gls{set-size-bound}/2\rfloor +1, 1]\)
        }
}

\newglossaryentry{set-size-bound}
{
    name=\ensuremath{ss_{\max}},
    description={
            Constraint on probabilistic set size
        }
}

\newglossaryentry{halting-condition-encapsulation}
{
    name=\ensuremath{hc},
    description={
            Halting condition encapsulation
        }
}

\newglossaryentry{unique-nodes-found-in-gathering}
{
    name=\ensuremath{\Gamma},
    description={
            Subset of network nodes encountered in a \gls{gathering}
        }
}
\newglossaryentry{unique-nodes-found-in-gathering-in-step}
{
name=\ensuremath{\gls{unique-nodes-found-in-gathering}_{\idx{0}}},
description={
        Subset of network nodes encountered up to step \idx{0}
    }
}

\newglossaryentry{next-draw-in-gathering}
{
    name=\ensuremath{nxtDraw},
    description={
            Identifier of a next \gls{draw} in a \gls{gathering}
        }
}

\newglossaryentry{peer-list-response}
{
    name=\ensuremath{\aleph},
    description={
            Response to a peer list request, a of node identifiers
        }
}

\newglossaryentry{population-size}
{
    name=\ensuremath{N},
    description={
            Population size parameter for the construction of a hypergeometric distribution
        }
}

\newglossaryentry{number-of-successes-in-population}
{
    name=\ensuremath{K},
    description={
            Number of success in the population
        }
}

\newglossaryentry{sample-size}
{
    name=\ensuremath{n},
    description={
            The sample size
        }
}
\newglossaryentry{number-of-successes-in-sample}
{
    name=\ensuremath{k},
    description={
            Number of success in the sample
        }
}

\newglossaryentry{network-size}
{
    name=\ensuremath{|\gls{network-nodes-set}|},
    description={
            Network size
        }
}
\newglossaryentry{network-nodes-set}
{
    name=\ensuremath{S},
    description={
            A set of all network nodes
        }
}
\newglossaryentry{number-of-malicious-nodes-in-the-network}
{
    name=\ensuremath{M},
    description={
            Number of malicious nodes in the network
        }
}

\newglossaryentry{network-topology-mapping}
{
    name=\ensuremath{T},
    description={
            Network topology mapping
        }
}
\newglossaryentry{network-topology-mapping-element}
{
    name=\ensuremath{t},
    description={
            A network topology mapping element
        }
}
\newglossaryentry{node-capacities}
{
    name=\ensuremath{\hat{c}},
    description={
            Node capacities vector
        }
}
\newglossaryentry{node-capacities-element}
{
    name=\ensuremath{c},
    description={
            Node capacities vector element
        }
}
\newglossaryentry{node-weights}
{
    name=\ensuremath{\hat{w}},
    description={
            Node weights vector
        }
}

\newglossaryentry{node-weights-element}
{
    name=\ensuremath{w},
    description={
            Node weights vector element
        }
}
\newglossaryentry{weights-unfamiliar-to-node}
{
    name=\ensuremath{\hat{U}},
    description={
            Vector of weights containing all weights of nodes unfamiliar to node \(\gls{network-topology-mapping-element}_{l}\)
        }
}
\newglossaryentry{number-of-draws}
{
    name=\ensuremath{d},
    description={
            Number of \glspl{draw} made in a \gls{gathering}
        }
}
\newglossaryentry{exponential-distribution-parameter}
{
    name=\ensuremath{\lambda},
    description={
            Exponential distribution parameter
        }
}
\newglossaryentry{unique-nodes-per-draw-random-variable}
{
    name=\ensuremath{Y},
    description={
            Random variable denoting the number of unique nodes found throughout a single \gls{gathering}
        }
}

\newglossaryentry{average-unique-nodes-per-draw-random-variable}
{
    name=\ensuremath{Z},
    description={
            Random variable denoting the average number of unique nodes found throughout a single \gls{gathering}, \(\gls{average-unique-nodes-per-draw-random-variable} \sim \mathcal{N}(\mu,\,\sigma^{2})\)
        }
}
\newglossaryentry{generic-random-variable}
{
    name=\ensuremath{X},
    description={
            Random variable
        }
}
\newglossaryentry{generic-random-variable-occurrence}
{
    name=\ensuremath{x},
    description={
            Random variable occurrence
        }
}
\newglossaryentry{maximum-execution-time}
{
    name=\ensuremath{t_{\max}},
    description={
            Maximum execution time in seconds
        }
}
\newglossaryentry{ratio-ps-sub-linear-function-of-desired-malicious-node-tolerance}
{
    name=\ensuremath{\omega},
    description={
            Measured ratio between \gls{amount-of-seen-nodes} and \gls{desired-malicious-node-tolerance} when the corresponding probabilistic sets size starts to behave as a sub-linear function of \gls{desired-malicious-node-tolerance}
        }
}

\newglossaryentry{pss}
{
    name=\ensuremath{\Pi_{s}},
    description={
            Probabilistic Safe Set
        }
}

\newglossaryentry{pps}
{
    name=\ensuremath{\Pi_{p}},
    description={
            Probabilistic Progress Set
        }
}
\newglossaryentry{constant}
{
    name=\ensuremath{C},
    description={
            A constant
        }
}
\newglossaryentry{new-node}
{
    name=\ensuremath{a},
    description={
            A new node entering a distributed ledger network
        }
}
\newglossaryentry{gathering}
{
    name=gathering,
    description={
            The process of exploring the network
        },
    plural={gatherings}
}
\newglossaryentry{draw}
{
    name=draw,
    description={
            A step performed during network exploration
        },
    plural={draws}
}

\begin{document}
%
% paper title
% Titles are generally capitalized except for words such as a, an, and, as,
% at, but, by, for, in, nor, of, on, or, the, to and up, which are usually
% not capitalized unless they are the first or last word of the title.
% Linebreaks \\ can be used within to get better formatting as desired.
% Do not put math or special symbols in the title.
\title{Aurora: a probabilistic algorithm for distributed ledgers enabling trustless synchronization and transaction inclusion verification}
%
%
% author names and IEEE memberships
% note positions of commas and nonbreaking spaces ( ~ ) LaTeX will not break
% a structure at a ~ so this keeps an author's name from being broken across
% two lines.
% use \thanks{} to gain access to the first footnote area
% a separate \thanks must be used for each paragraph as LaTeX2e's \thanks
% was not built to handle multiple paragraphs
%
%
%\IEEEcompsocitemizethanks is a special \thanks that produces the bulleted
% lists the Computer Society journals use for "first footnote" author
% affiliations. Use \IEEEcompsocthanksitem which works much like \item
% for each affiliation group. When not in compsoc mode,
% \IEEEcompsocitemizethanks becomes like \thanks and
% \IEEEcompsocthanksitem becomes a line break with idention. This
% facilitates dual compilation, although admittedly the differences in the
% desired content of \author between the different types of papers makes a
% one-size-fits-all approach a daunting prospect. For instance, compsoc 
% journal papers have the author affiliations above the "Manuscript
% received ..."  text while in non-compsoc journals this is reversed. Sigh.

\author{Federico~Matteo~Benčić,~\IEEEmembership{Member,~IEEE,}
  Ivana~Podnar~Žarko,~\IEEEmembership{Member,~IEEE,}
  % \thanks{}
  %\thanks{Manuscript received April 19, 2005; revised August 26, 2015.}
  }

% note the % following the last \IEEEmembership and also \thanks - 
% these prevent an unwanted space from occurring between the last author name
% and the end of the author line. i.e., if you had this:
% 
% \author{....lastname \thanks{...} \thanks{...} }
%                     ^------------^------------^----Do not want these spaces!
%
% a space would be appended to the last name and could cause every name on that
% line to be shifted left slightly. This is one of those "LaTeX things". For
% instance, "\textbf{A} \textbf{B}" will typeset as "A B" not "AB". To get
% "AB" then you have to do: "\textbf{A}\textbf{B}"
% \thanks is no different in this regard, so shield the last } of each \thanks
% that ends a line with a % and do not let a space in before the next \thanks.
% Spaces after \IEEEmembership other than the last one are OK (and needed) as
% you are supposed to have spaces between the names. For what it is worth,
% this is a minor point as most people would not even notice if the said evil
% space somehow managed to creep in.

% The paper headers
% The paper headers have been removed prior to submission, return them lines 1060-1063
\markboth{}%
% \markboth{Journal of \LaTeX\ Class Files,~Vol.~14, No.~8, August~2015}%
{}
\IEEEtitleabstractindextext{%
  \begin{abstract}
    A new node joining a blockchain network first synchronizes with the network to verify ledger state by downloading the entire ledger history.
    We present Aurora, a probabilistic algorithm that \textit{identifies honest nodes} for transient or persistent communication in the presence of malicious nodes in a blockchain network, or ceases operation if it is unable to do so. The algorithm allows a node joining the network to make an informed decision about its next synchronization step or to verify that a transaction is contained in a valid ledger block without downloading the entire ledger or even the header chain.
    The algorithm constructs a Directed Acyclic Graph on the network topology to select a subset of nodes including a predefined number of honest nodes with a given probability.
    It is evaluated on a Bitcoin-like network topology using an open-source blockchain simulator. We investigate algorithm performance and analyze its communication complexity. Our results show that the algorithm facilitates trustless interactions of resource-constrained nodes with a blockchain network containing malicious nodes to enable a leaner initial blockchain download or an efficient and trustless transaction inclusion verification. Moreover, the algorithm can be implemented without any changes to the existing consensus protocol.
  \end{abstract}

  % Note that keywords are not normally used for peerreview papers.
  }

% make the title area
\maketitle

% To allow for easy dual compilation without having to reenter the
% abstract/keywords data, the \IEEEtitleabstractindextext text will
% not be used in maketitle, but will appear (i.e., to be "transported")
% here as \IEEEdisplaynontitleabstractindextext when compsoc mode
% is not selected <OR> if conference mode is selected - because compsoc
% conference papers position the abstract like regular (non-compsoc)
% papers do!
\IEEEdisplaynontitleabstractindextext %chktex 1
% \IEEEdisplaynontitleabstractindextext has no effect when using
% compsoc under a non-conference mode.

% For peer review papers, you can put extra information on the cover
% page as needed:
% \ifCLASSOPTIONpeerreview
% \begin{center} \bfseries EDICS Category: 3-BBND \end{center}
% \fi
%
% For peerreview papers, this IEEEtran command inserts a page break and
% creates the second title. It will be ignored for other modes.
\IEEEpeerreviewmaketitle %chktex 1

\ifCLASSOPTIONcompsoc %chktex 1
  \IEEEraisesectionheading{\section{Introduction}\label{sec:introduction}}
\else
  \section{Introduction}\label{sec:introduction}
\fi
% Computer Society journal (but not conference!) papers do something unusual
% with the very first section heading (almost always called "Introduction").
% They place it ABOVE the main text! IEEEtran.cls does not automatically do
% this for you, but you can achieve this effect with the provided
% \IEEEraisesectionheading{} command. Note the need to keep any \label that
% is to refer to the section immediately after \section in the above as
% \IEEEraisesectionheading puts \section within a raised box.

% The very first letter is a 2 line initial drop letter followed
% by the rest of the first word in caps (small caps for compsoc).
% 
% form to use if the first word consists of a single letter:
% \IEEEPARstart{A}{demo} file is ....
% 
% form to use if you need the single drop letter followed by
% normal text (unknown if ever used by the IEEE):
% \IEEEPARstart{A}{}demo file is ....
% 
% Some journals put the first two words in caps:
% \IEEEPARstart{T}{his demo} file is ....
% 
% Here we have the typical use of a "T" for an initial drop letter
% and "HIS" in caps to complete the first word.
% Blockchain - short overview
\IEEEPARstart{D}{istributed Ledger} Technology (DLT) is a revolutionary technology for digitizing assets~\cite{ghosh2019blockchain} and has become widely known through one of its specializations, the blockchain. Blockchain is a distributed ledger based on a peer-to-peer (P2P) network to manage transactions for multiple entities in a verifiable and traceable manner. Data is written within blocks of blockchain transactions, and does not require a centralized entity, but rather the entire P2P network, to maintain the blocks. The main features of blockchain are immutability, transparency and fault tolerance~\cite{zheng2019survey}.

% Blockchain - application domains - excluded for space
% The application of blockchain has been considered in a variety of domains. After its inception as a cryptocurrency, it has been explored as a tool for signing smart contracts, in communication systems,in healthcare, in the Internet of Things (IoT), in financial systems, and in electronic voting, to name a few~\cite{saad2020exploring}. In IoT, as one of the promising application domains, blockchain can be used to improve interoperability across IoT devices, traceability and reliability of IoT data, as well as facilitate autonomous interaction of IoT devices~\cite{dai2019blockchain}.

% Different types of nodes
Users running blockchain nodes can choose between two types of nodes: full and light. Full nodes download and verify the entire ledger which contains all transactions since the ledger creation. Consequently, they operate in a trustless manner, but also have more stringent hardware requirements. In contrast, light nodes only need to download part of the ledger (e.g., the header chain) and therefore consume less processing power, network bandwidth and memory compared to full nodes. However, light nodes depend on full nodes --- full nodes provide light nodes with the metadata required for their operation.

% The size of the blockchain is immense
Regardless of the node type and due to the immutability requirement, the size of the ledger is immense as the number of transactions continuously increases. This leads to a significant load during node synchronization. For example, in January 2020, the Ethereum blockchain had about 250 GB in non-archived mode~\cite{gabizon2020plumo}. Bitcoin's header chain had 50 MB in April 2020, while Ethereum's header chain had 5 GB~\cite{zamyatin2020txchain} and is growing about 1 GB per year~\cite{lu2020generic}. Since the synchronization process is memory and time consuming, a new node may either be unable to download the ledger or unwilling to wait too long for the process to complete. For example, nodes in the Internet of Things (IoT) domain are often resource-constrained devices with limited memory, processing power, and energy, which further exacerbates the problem. As a result, users often opt for third-party  explorers instead of running their own nodes.

% Initial blockchain synchronization is centralized
In addition to high resource consumption, the process of node synchronization (i.e., bootstrapping) becomes a point of centralization, as it depends in part on a set of well known nodes that are assumed to be both available and honest. Today's client implementations rely on a list of well known addresses~\cite{homoliak2019security} that may not always be available or may even exhibit Byzantine behavior. One could work around the issue of unavailable seeds by manually adding bootstrap peers, but manually added nodes may also exhibit malicious behavior, making a new node vulnerable to various types of attacks, from denial of service to Eclipse attacks~\cite{heilman2015eclipse, saad2020exploring}. The consequences of a malicious node being a first contact node can vary for a victim. They range from a waste of time and resources, in the best case, to a state where, in the worst case, the victim is unaware of the existence of a longer chain because no honest node is available to advertise it.
% Partitioned nodes have the same effect as malicious nodes.
Note that if a new node joins the network through a set of partitioned nodes (which may or may not be malicious), the new node could synchronize with a ledger that is in a state of extended fork, and such a state can lead to a double-spending attack. For simplicity, hereinafter we do not distinguish between partitioned and malicious Byzantine nodes, as our algorithm is applicable in both scenarios.

% Aurora as the solution to the problems listed above
The Aurora algorithm, originally proposed in~\cite{benvcic2019aurora}, is a consensus-agnostic probabilistic algorithm designed to enable a new node to avoid the adverse influence of Byzantine or malicious nodes in ledger networks by detecting a predefined number of honest nodes with a high and controllable probability. The identified nodes are used thereafter for persistent or transient communication of the new node with the network. If the predefined number of honest nodes cannot be identified, the algorithm stops.
Assuming that an honest node reveals to a new node what is accepted as canonical truth by the majority of the voting power, as opposed to a malicious node that tries to subvert the new node, the algorithm saves the node's resources by efficiently discovering \textit{a subset of network nodes} containing at least \(1\) or \(1+|\gls{number-of-malicious-nodes-in-the-network}|\)  honest nodes, where \gls{number-of-malicious-nodes-in-the-network} is the set of malicious network nodes. Such subsets of network nodes can be used for trustless ledger synchronization, or transaction inclusion verification without the need to download the entire ledger or even the header chain.

% Advantages of the use of Aurora 
While the algorithm can be used by both full and light nodes, its low memory footprint and stochastic behavior make it particularly suitable for resource-constrained devices, where it offers the greatest benefit. Although potentially applicable in other domains, the algorithm offers particular benefits in the domain of public and permissionless DLT solutions, as it reinforces the decentralized and trustless ethos on which the technology is based.

% Difference from our previous work and contributions.
This work builds on our previous work which introduced the initial version of the Aurora algorithm~\cite{benvcic2019aurora}. The main contributions of the paper and the most notable distinctions compared to our previous work are the following:

\begin{enumerate}
  \item Redefinition of the algorithm to provide an accurate probabilistic output and communication complexity of the algorithm. The algorithm identifies a subset of network nodes containing a predefined number of honest nodes with high probability as a function of the assumed number of malicious network nodes.
  \item A comprehensive evaluation of the redefined algorithm is performed by simulation on a realistic network topology using open-source simulation tools in a Bitcoin-like network environment. The simulation results confirm our analytical findings.
  \item Two concrete application examples of the algorithm:
        \begin{enumerate}
          \item In a scenario where a new node joins a distributed ledger network, the algorithm assists the node during the synchronization process with a blockchain network with malicious nodes.
          \item The algorithm facilitates trustless and efficient verification of transaction inclusion in a block with a predefined correctness probability and significantly reduces the resource consumption of a device executing the verification procedure.
        \end{enumerate}
\end{enumerate}

% Remark about the resemblances with other P2P
Although our work is relevant to solutions for peer-to-peer (P2P) networks which detect malicious actors or counter their direct influence, our work differs from such solutions in that it does not detect malicious nodes, but rather identifies node sets containing honest nodes. Moreover, our work has a well-defined application domain, namely DLT networks. Comparison with related work is discussed in more detail in~\cref{sec:related-work}.

% Structure of the Paper.
The paper is organized as follows:~\cref{sec:related-work} compares our solution with other existing solutions.~\cref{sec:preliminaries} introduces necessary terms and definitions, and enumerates the assumptions under which our algorithm works.~\cref{sec:probabilistic-set-size} discusses how relevant variables affect the size of sets that contain a predefined number of honest nodes.~\cref{sec:collecting-a-subset-of-the-network} presents the Aurora algorithm and introduces two relevant applications of the algorithm, while~\cref{sec:communication-complexity} investigates its communication complexity.~\cref{sec:methodology-and-experiments} explains the methodology used for algorithm verification, the experiments performed, and their results.~\cref{sec:challenges-and-future-work} discusses possible challenges and future work, while~\cref{sec:conclusion} concludes the paper with a glossary appended at the end of the paper.

\section{Related work}~\label{sec:related-work}
% Since more people are like you (we have 1500 publicly routed nodes vs. 9000 total nodes), there's a slot burden on the public nodes.
% https://github.com/ethereum/go-ethereum/issues/1563
% For geth, the ratio is 50%. If you specify --maxpeers 25, it will make 12 connections and leave the rest of the slots for inbound connections.
% --maxpeers value Maximum number of network peers (network disabled if set to 0) (default: 50)
% https://github.com/ethereum/go-ethereum/issues/1563
% Structure of the section
We begin by taking a closer look at light-client solutions, highlighting potential synergies and contrasting differences to our algorithm. Next, since our work is the most valuable when adversarial influences are present in both P2P and DLT networks, we continue with a brief overview of relevant works.

Light client implementations fall broadly into the following three categories: remote clients, simplified payment verification clients, and trustless clients.
% FlyClient
% Nipopow
% Mimblewimble https://youtu.be/aHTRlbCaUyM?t=752 - needs header chain
%https://youtu.be/aHTRlbCaUyM?t=1930 - compression when bootstrapping
% BTC SPV
% Vault: Fast Bootstrapping for the Algorand Cryptocurrency - https://www.youtube.com/watch?v=3HLpsE3LeAY
% Remote clients.

\textbf{Remote clients} rely on a centralized solution for communication with a DLT network. An example is MetaMask, an Ethereum wallet that communicates with the Ethereum ledger via the Infura Gateway System~\cite{lee2019beginning}. Remote clients which are not completely centralized also exist, for example the SlockIt INCUBED client\footnote{https://consensys.net/diligence/audits/2019/09/slock.it-incubed3/}. This client uses multiple gateways to communicate with a ledger network. The gateways have a stake in the network that can be reduced if their misbehavior is discovered. Such clients sacrifice the integrity of the data by trusting a third party in exchange for a very small computational resource requirement to interact with a DLT network. They differ greatly from our solution because our solution is completely decentralized.

% Simplified Payment Verification clients (SPVs).
% Mimblewimble https://youtu.be/aHTRlbCaUyM?t=745 - needs header chain
%https://youtu.be/3HLpsE3LeAY?t=598
\textbf{Simplified Payment Verification clients (SPVs)} are standard distributed ledger clients that synchronize with the network's header chain and request the rest of the block information as needed. Examples are Electrum\footnote{https://electrum.readthedocs.io/en/latest/faq.html} for Bitcoin, or Geth (in light mode) for Ethereum. There are SPVs that, similar to our solution, recognize the added value of multiple sources of truth and can even be used in conjunction with our solution. For example, Electrum maintains connections to approximately \(10\) servers and subscribes to block header notifications to all of them to detect forks and partitions. Here we see an emerging synergy: The Aurora algorithm can be used to identify \(10\) servers which are honest with high probability. The Tendermint~\cite{kwon2014tendermint} light client acquires prerequisites for connecting to the network by means outside of Tendermint\footnote{https://docs.tendermint.com/master/spec/p2p/node.html} (e.g., social consensus), and thus it can apply our solution to identify these prerequisites. SPVs differ from remote clients since they maintain their trustlessness to some extent by validating metadata against the header chain in their possession. However, since they are served by full nodes, they rely on full nodes which are assumed to be available and honest. Moreover, the applicability of SPVs to resource-constrained devices is debatable --- even the header chain may be too large for some devices. Our solution differs greatly from SPVs since it does not require the header chain to work.

% Trustless clients
\textbf{Trustless clients} attempt to maintain the trustless mechanisms of SPVs while keeping their size sublinear or constant compared to ledger size or even header chain size. As such, they are sometimes referred to as \emph{super light clients}. A client running the Aurora algorithm would partially fall into this category --- when used for transaction inclusion verification, the client does not need to download the header chain (or part of the header chain). A proposal from~\cite{xu2018efficient} uses a cryptography accumulator and generates a chain summary in the form of a block attribute. Then, the client must randomly choose a slice of the network nodes, and if the majority of nodes is compromised by the attacker, the protocol is not secure. Consequently, our algorithm can be used in conjunction with the above --- the slice can be the output of our algorithm. Vault~\cite{leung2019vault} is a solution built on a proof-of-stake consensus protocol proposed by Algorand~\cite{chen2016algorand} and introduces a fast bootstrapping method relying on the presence of stamping certificates, while our solution does not require any additional data structures to operate.

% Explain why FlyClient and BlockQuick are highlighted.
We continue by highlighting two solutions: FlyClient~\cite{bunz2020flyclient} as an example of a solution that would benefit significantly from our algorithm, and BlockQuick~\cite{letz2019blockquick} as an example of a solution that addresses the potential existence of Eclipse attacks, but does so in a fundamentally different way than our solution.

% https://www.youtube.com/watch?v=E785K59FaxU
% FlyClient summary.
\textbf{FlyClient} is a light client proposal for Proof of Work blockchains~\cite{bunz2020flyclient}. The client only needs to store a logarithmic number of block headers to provide strong mechanisms for ensuring the validity of those block headers by using techniques such as probabilistic sampling, MRRs\footnote{Merkle Mountain Ranges}, and the Fiat-Shamir heuristic. As FlyClient requires nodes to maintain MRRs, FlyClient cannot be used on the currently running Bitcoin and Ethereum networks without forks. Also, the client must be connected to at least one honest node, which means that FlyClient cannot protect a node against Eclipse attacks, unlike the Aurora client. Thus, our client does not compete with this solution, but can work in synergy with it. The one honest node that FlyClient needs to operate can be found with our solution.

% BlockQuick summary.
\textbf{BlockQuick} is based on a consensus-based reputation scheme~\cite{letz2019blockquick}. A BlockQuick client maintains the Consensus Reputation Table, which contains miner addresses with the highest reputation shares generated based on the latest \(100\) block headers. The reputation share of a miner is equal to the number of blocks mined in the last \(100\) blocks. When new blocks are broadcasted, the client validates the miner's cryptographic signature against the data from the Consensus Reputation Table. A new block is considered valid only if the block receives a consensus share score \(>50\%\). Thus, it is not just a matter of choosing the longest chain with the highest difficulty, but accepting the block from miners with a high consensus share. The client is resistant to both MITM and Eclipse attacks.
% Changes required to make BlockQuick work 
Similar to FlyClient, the requirements for running BlockQuick clients with the current Ethereum and Bitcoin networks are not met. According to~\cite{letz2019blockquick}, each miner should be reachable at the address specified in the block, each block header must contain an address and the public ID of the block miner and each block must contain a proof of inclusion of all previous block headers. The result is that, unlike our solution, a BlockQuick client cannot be deployed on currently running Ethereum or Bitcoin networks without a hard fork.

% Adversarial influence in P2P networks.
Relevant solutions in the P2P domain are anomaly detection methods and sibyl group inference solutions (e.g., SybilGuard~\cite{yu2006sybilguard}). However, since reputation measurements are not available in existing blockchain solutions, such approaches are not applicable in this context. Anomaly detection techniques~\cite{ding2012intrusion}, flow-based and graph-based methods~\cite{chen2013ascos,chowdhury2017botnet} as well as network traffic reliant solutions (e.g.,~\cite{zhang2012network,zhang2013effective}) differ from our solution in two key aspects: The first and most obvious is the application domain --- our work is DLT-specific. Second, we do not deal with or infer the identity of malicious clusters, but focus on identifying honest nodes for bootstrapping and transaction inclusion verification.

% Adversarial influence in DLT networks.
As for the adversarial influence in DLT networks, previous studies, e.g.,~\cite{wust2016ethereum,marcus2018low,xu2020eclipsed,heilman2015eclipse,homoliak2019security,apostolaki2017hijacking, al2018fraud}, have confirmed that prominent DLT solutions such as Bitcoin~\cite{nakamoto2019bitcoin} and Ethereum~\cite{wust2016ethereum} are vulnerable to Denial of Service, Eclipse Attacks, BGP highjacks, Man-in-the-Middle (MITM) attacks, network partitioning, etc. The consequences of adversarial influences are non-trivial and range from transaction dropping to double spending. Although the related work proposes and implements several countermeasures, it differs from our work in two key ways. First, we do not directly counteract malicious clusters, but focus on detecting honest nodes for persistent or transient communication, thus circumventing the malicious influence rather than countering it or dealing with the consequences. Second, our algorithm does not require special data structures or protocol modifications to be integrated with deployed DLT solutions.

\section{Preliminaries}\label{sec:preliminaries}
% The scenario from which we derive all terms and definitions.
Let us consider a network \gls{network-nodes-set} with \gls{number-of-malicious-nodes-in-the-network} malicious nodes. We assume the following:

% We found a random node on the Internet
\begin{assumption}\label{con:no-bootstrap-peers}
  A new node \(\gls{new-node} \in \gls{network-nodes-set}\) joins the network by contacting a single first-contact node \gls{first-contact} and is unaware of any other network node (e.g., bootstrap peers are not available).
\end{assumption}
% We don't know whether \gls{first-contact} is malicious or not.
\begin{assumption}
  The first-contact node \gls{first-contact} may be malicious, and may collude with other malicious peers to subvert node \gls{new-node}.
\end{assumption}
% We can try to explore the network via \gls{first-contact}
\begin{assumption}
  A subset of nodes from \gls{network-nodes-set}, denoted by \gls{unique-nodes-found-in-gathering}, contains nodes discoverable by node \gls{new-node} via \gls{first-contact}.
\end{assumption}

\noindent By \textit{discoverable nodes} we refer to those nodes that node \gls{new-node} has encountered or become aware of their existence when its first-contact node is \gls{first-contact}.
% We can collect at least \gls{desired-malicious-node-tolerance} + 1 node. 
\begin{assumption}\label{con:at-least-one-honest-node}
  There is a set of malicious nodes, \(\gls{number-of-malicious-nodes-in-the-network} \subset \gls{network-nodes-set}\), the size of \gls{number-of-malicious-nodes-in-the-network} is \gls{desired-malicious-node-tolerance} and the relation \(\gls{desired-malicious-node-tolerance} <\gls{amount-of-seen-nodes}\) holds. The value of \gls{desired-malicious-node-tolerance} is known a priori or can be inferred (the initialization of \gls{desired-malicious-node-tolerance} is discussed in the following sections).
\end{assumption}
% A minority of adversarial influence.
\begin{assumption}\label{con:honest-majority}
  The adversary always holds a minority of the voting power.
\end{assumption}
% Number of malicious nodes can not change during execution.
\begin{assumption}\label{con:malicious-nodes-are-constant}
  The number of malicious nodes in the network cannot increase during algorithm execution.
\end{assumption}

% Algorithm gist
Under these assumptions, node \gls{new-node} performs a process of network exploration which we name \textit{\gls{gathering}}, as the process of collecting discoverable nodes initiated from node \gls{first-contact} to create \gls{unique-nodes-found-in-gathering}. Each node queried during a \gls{gathering} exposes a set of its neighboring nodes to node \gls{new-node}, which \gls{new-node} uses to expand \gls{unique-nodes-found-in-gathering}. We continue by providing definitions based on the previously stated assumptions.

% What is a deterministic set?
\begin{definition}
  [Deterministic Honest Set, \gls{dsh}]
  For a given \gls{unique-nodes-found-in-gathering}, a set \gls{dsh} is a subset of \gls{unique-nodes-found-in-gathering} which contains at least \idx{2} honest nodes.
\end{definition}

% We can always find at least one honest node. 
\begin{corollary}\label{def:corollary-ds1-always-exists}
  For any given \gls{unique-nodes-found-in-gathering}, there exists a deterministic honest set with at least 1 honest node, \(\gls{ds}_{1}\).

  \textit{Proof}. Iff the relation \(\gls{desired-malicious-node-tolerance} <\gls{amount-of-seen-nodes}\) holds as per~\cref{con:at-least-one-honest-node}, \gls{unique-nodes-found-in-gathering} must contain at least one honest node. Consequently, \(\gls{ds}_{1}\) can always be constructed.
\end{corollary}

% What is a probabilistic set?
\begin{definition}
  [Probabilistic Honest Set, \gls{psh}]
  For a given \gls{unique-nodes-found-in-gathering}, a set \gls{psh} is a subset of \gls{unique-nodes-found-in-gathering} which contains at least \idx{2} honest nodes with probability \gls{correctness-probability}. The probability \gls{correctness-probability} is derived from an underlying hypergeometric distribution.
\end{definition}

% Further restrictions for \gls{psh} with an explanation of the restrictions.
To reduce problem complexity, we introduce justifiable constraints on \idx{2} relevant to the DLT domain. First, we consider those \gls{dsh} and \gls{psh} that contain at least one honest node, which guarantees that the truth (e.g., the longest blockchain) can eventually be received by node \gls{new-node}. Second, we consider those \gls{dsh} and \gls{psh} that contain a majority of honest nodes, which guarantees that an honest answer can be derived by majority voting without requiring a resource-intensive process of ledger/header chain analysis. With respect to the above constraints on \idx{2}, we define the corresponding deterministic sets that will be used as a naive baseline as follows:

% What is a deterministic safe set?
% Sigurno ima barem 1 iskren čvor jer ima k+1, k u bizanstkom ispadu
\begin{definition}
  [\gls{deterministic-safe-set}, \(\gls{dss} \equiv \gls{ds}_{1}\)]

  For a given \gls{unique-nodes-found-in-gathering} containing \gls{desired-malicious-node-tolerance} malicious nodes, where \(\gls{amount-of-seen-nodes} > \gls{desired-malicious-node-tolerance}\), \gls{dss} is a deterministic honest set which contains at least one honest node and its size is at least \(\gls{desired-malicious-node-tolerance}+1\).
\end{definition}

% Origin of the name.
We call this set \textit{safe set} since node \gls{new-node} can obtain at least one honest answer in the worst case by querying all members of \gls{dss}. By analyzing the obtained ledgers and verifying their transactions since genesis, the node can detect possible discrepancies in the ledger state as reported by the queried nodes, but also identify a correct ledger.

% What is a deterministic progress set?
% Sigurno sadrži 2k+1 čvorova, ima iskrenu većinu, dovoljno za sporazum
\begin{definition}
  [\gls{deterministic-progress-set}, \(\gls{dps} \equiv \gls{ds}_{\lfloor |\gls{ds}| / 2 \rfloor + 1}\)]

  For a given \gls{unique-nodes-found-in-gathering} containing \gls{desired-malicious-node-tolerance} malicious nodes, where \(\gls{amount-of-seen-nodes} > 2\gls{desired-malicious-node-tolerance}\), \gls{dps} is a deterministic honest set which contains a majority of honest nodes and its size is at least \(2\gls{desired-malicious-node-tolerance}+1\).
\end{definition}
% Origin of the name.
We call this set \textit{progress set} because node \gls{new-node} can obtain a majority of honest answers in the worst case by querying all members of the set and perform an action based on the majority vote about the ledger state reported by the queried nodes.

% Explain benchmark.
The deterministic baselines, both safe and progress sets, offer a naive solution to the problem of identifying a truthful node: A node can trivially query multiple network nodes under the assumption that there are at most \gls{desired-malicious-node-tolerance} malicious nodes, and compare the obtained answers in search of the truth. However, this method is very inefficient, as we show in~\cref{sec:probabilistic-set-size}.
Let us now consider definitions of the corresponding sets with probabilistic bounds.

% What is a probabilistic safe set?
% Ima barem 1 iskren čvor uz vjerojatnost rho, k u bizanstkom ispadu
\begin{definition}
  [\gls{probabilistic-safe-set}, \(\gls{pss} \equiv \gls{ps}_{1}\)]\label{def:pss}

  For a given \gls{unique-nodes-found-in-gathering} containing \gls{desired-malicious-node-tolerance} malicious nodes, where \(\gls{amount-of-seen-nodes} > \gls{desired-malicious-node-tolerance}\), \gls{pss} is a probabilistic honest set which contains at least one honest node with probability \gls{correctness-probability}.
\end{definition}

% What is a probabilistic progress set?
% Sadrži 2k+1 čvorova uz vjerojatnost rho, ima iskrenu većinu, dovoljno za sporazum
\begin{definition}
  [\gls{probabilistic-progress-set}, \(\gls{pps} \equiv \gls{ps}_{\lfloor |\gls{ps}| /2  \rfloor + 1}\)]\label{def:pps}

  For a given \gls{unique-nodes-found-in-gathering} containing \gls{desired-malicious-node-tolerance} malicious nodes, where \(\gls{amount-of-seen-nodes} > 2\gls{desired-malicious-node-tolerance}\), \gls{pps} is a probabilistic honest set which contains a majority of honest nodes with probability \gls{correctness-probability}.
\end{definition}

% Explanation of the baseline choice.
%In the following we explain the process of constructing probabilistic sets, which is an iterative endeavour.

%\subsection{How to construct a probabilistic set?}
% Explain that construction is iterative
Probabilistic sets are generated during an iterative process when node \gls{new-node} in each step \(1)\) contacts a network node, \(2)\) asks for a set of its neighbors, and then \(3)\) selects a next-draw node to repeat the process. For this reason, we define the following: %chktex 10

% The set from which the nodes are drawn in step \idx{1}.
\begin{definition}
  [Set of discoverable network nodes in step \idx{0}, \gls{unique-nodes-found-in-gathering-in-step}]\label{def:unique-nodes-found-in-gathering-in-step}

  \gls{unique-nodes-found-in-gathering-in-step} is a subset of network nodes which node \gls{new-node} has learned about while performing a \gls{gathering} up to step \idx{0}, where \(|\gls{unique-nodes-found-in-gathering-in-step}| \leq |\gls{unique-nodes-found-in-gathering}_{\idx{0}+1}|\).
\end{definition}

% What is a hypergeometric distribution?
In other words, \gls{unique-nodes-found-in-gathering-in-step} is filled by nodes appearing in neighbor sets (i.e., peer lists) reported by contacted nodes, and this set contains candidates for probabilistic honest sets. \gls{psh} is generated from a finite population \gls{unique-nodes-found-in-gathering-in-step} by sampling random nodes without replacement. The sampling process can be modeled by the hypergeometric distribution~\cite{rivadulla1991mathematical}, as each element selected from \gls{unique-nodes-found-in-gathering-in-step} can be classified as a failure or a success. In our context, a success denotes the selection of an honest node, while a failure occurs when a malicious node is selected. In summary, the distribution models the probabilities associated with the number of successes in a hypergeometric experiment, and is defined by

\begin{equation}
  \label{eqn:hypergeometric-distribution-definiton}
  \gls{generic-random-variable} \sim Hypergeometric(\gls{population-size},\gls{number-of-successes-in-population},\gls{sample-size}),
\end{equation}

\noindent where \gls{population-size} is the population size, \gls{number-of-successes-in-population} is the number of successes in the population, and \gls{sample-size} is the sample size.

% Hypergeometric probability mass function.
The underlying probability mass function is then given by

\begin{equation}
  \label{eqn:hypergeometric-probability-mass-function}
  P(\gls{generic-random-variable}=\gls{generic-random-variable-occurrence}) =
  \frac{\binom{\gls{number-of-successes-in-population}}{\gls{number-of-successes-in-sample}}\binom{\gls{population-size}-\gls{number-of-successes-in-population}}{\gls{sample-size}-\gls{number-of-successes-in-sample}}}{\binom{\gls{population-size}}{\gls{sample-size}}} = f(\gls{population-size},\gls{number-of-successes-in-population},\gls{sample-size},\gls{number-of-successes-in-sample}),
\end{equation}

\noindent where \gls{number-of-successes-in-sample} is the number of successes in the sample.

% *** USEFUL BOOKMARKS ***
% Population sample needs to be drawn at random:
% 6. Monte Carlo Simulation-OgO1gpXSUzU @ 04:00
% Quote:
% "And the key fact that makes them work is that if we choose the sample at random, the sample will tend to exhibit the same properties as the population from which it is drawn."

% We should calculate the mean of the means
% 7. Confidence Intervals-rUxP7TM8-wo @ 21:35
% Quote:
% "And the reason is because we're not reasoning about a single spin of the wheel but about the mean of a set of spins."

% Construction of probabilistic sets explained casually.
We begin the construction of a probabilistic set by a \gls{gathering} initiated at node \gls{first-contact} to explore the network until \(|\gls{unique-nodes-found-in-gathering-in-step}|>\gls{desired-malicious-node-tolerance}\), and then start performing hypergeometric experiments for a given \gls{unique-nodes-found-in-gathering-in-step} (the population). It contains \(|\gls{unique-nodes-found-in-gathering-in-step}|-\gls{desired-malicious-node-tolerance}\) honest nodes (successes), and there exists a subset of size \(|\gls{psh}|\) (the sample) which contains \idx{2} honest nodes with probability \gls{correctness-probability} or more. More formally, the hypergeometric distribution can be stated as

\begin{equation}
  \label{eqn:hypergeometric-distribution-definiton-as-a-network-iteration}
  \gls{generic-random-variable} \sim Hypergeometric(
  |\gls{unique-nodes-found-in-gathering-in-step}|,
  |\gls{unique-nodes-found-in-gathering-in-step}|-\gls{desired-malicious-node-tolerance},
  |\gls{psh}|
  ),
\end{equation}

\noindent where we evaluate the following predicate
\begin{equation}
  \label{eqn:hypergeometric-distribution-predicate}
  p(X \geq \idx{2}) \geq \gls{correctness-probability}.
\end{equation}

% About the predicate
If the predicate is evaluated as true, we have constructed a probabilistic set of size \(|\gls{psh}|\) containing at least \idx{2} honest nodes with at least probability \gls{correctness-probability}. If the predicate is evaluated as false, we continue exploring the network.

% The pseudocode of the above casual explanation.
The pseudocode defining the construction of a probabilistic set is given in~\cref{alg:probabilistic-set-construction-pseudocode} which requires three input parameters: the correctness probability \gls{correctness-probability}, the first contact node \gls{first-contact}, and \gls{desired-malicious-node-tolerance}, the initialization of which is covered in~\cref{sec:variable-initialization-and-use-cases}.

\begin{algorithm}[ht!]
  \Input{
    \gls{correctness-probability} --- Probability guarantee
  }
  \Input{
    \(|\gls{psh}|\) --- Desired probabilistic set size
  }
  \Input{
    \(\gls{desired-malicious-node-tolerance}\) --- Desired malicious node tolerance
  }

  Gather at least \gls{desired-malicious-node-tolerance} + 1 nodes in \gls{unique-nodes-found-in-gathering}\;

  \(constructed \gets false\)\;

  \Do{\(!constructed\)}{
    \(
    \gls{generic-random-variable} \sim Hypergeometric(
    |\gls{unique-nodes-found-in-gathering}|,
    |\gls{unique-nodes-found-in-gathering}|-\gls{desired-malicious-node-tolerance},
    |\gls{psh}|
    )
    \)\;

    \uIf{\(p(X \geq \idx{2}) \geq \gls{correctness-probability} \)}{
      \(constructed \gets true\)\;
    }
    \Else{
      Add more nodes to \gls{unique-nodes-found-in-gathering}\;
    }

  }
  \caption{\(\gls{psh}\) construction}\label{alg:probabilistic-set-construction-pseudocode}
\end{algorithm}

% Outro.
Now that we understand how \(\gls{psh}\) is constructed, we can analyze why using probabilistic sets is advantageous compared to using their deterministic counterparts.

%\subsection{Why should we trade determinism for a probabilistic approach?}

\begin{figure}[ht!]
  \centering
  \includegraphics[width=\linewidth]{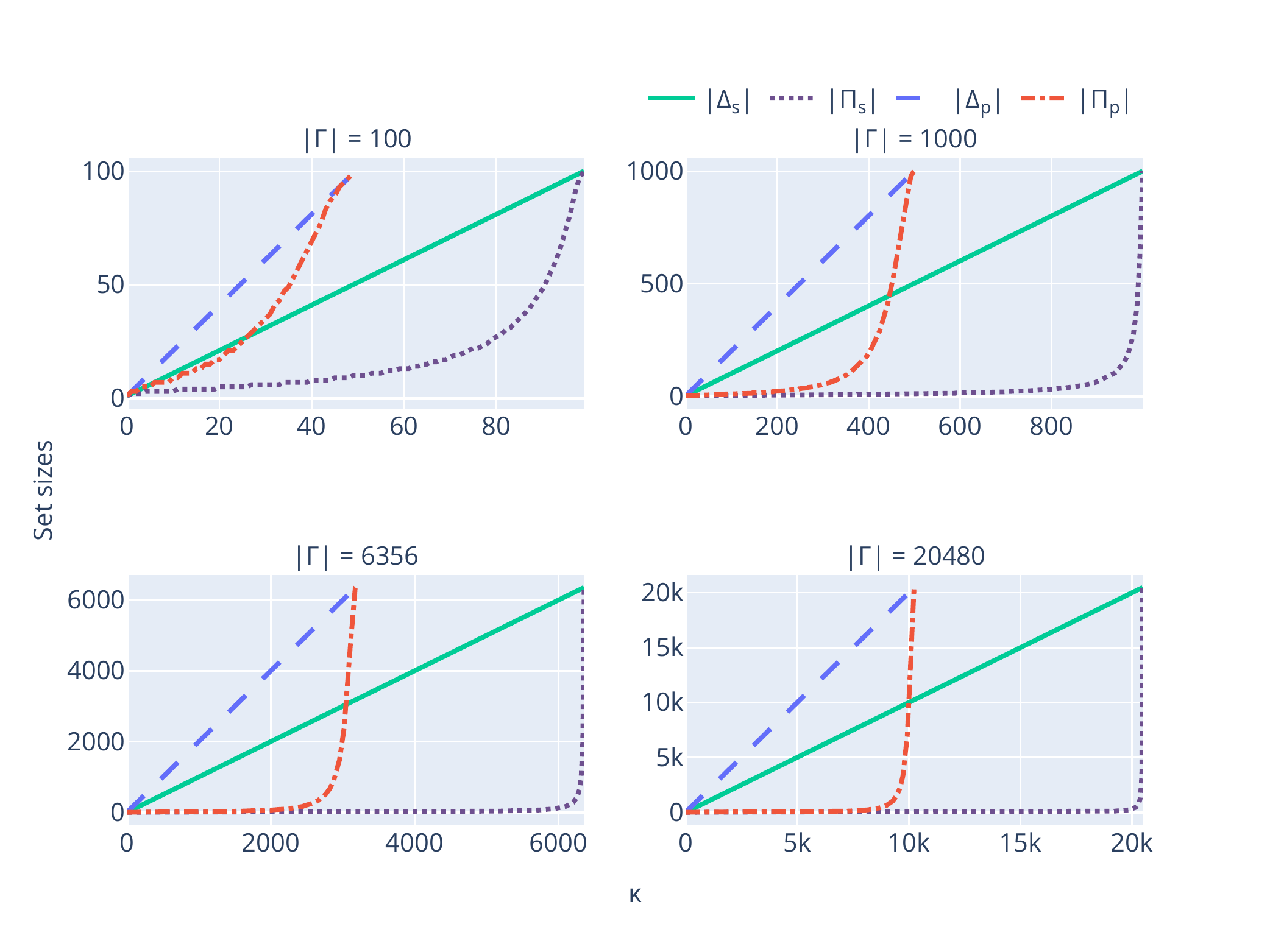}%
  \caption{Deterministic vs.\ probabilistic set sizes with an increase of \gls{desired-malicious-node-tolerance} for four different population sizes. When population size increases, we observe a greater difference between a probabilistic set size and its deterministic counterpart.}\label{fig:deterministic-vs-probabilistic-set-sizes-with-the-increase-of-malicious-nodes}
\end{figure}

%Explanation of experiments
We perform an experiment to compare the deterministic and probabilistic honest sets which are the most relevant to be used in the DLT domain, namely safe sets and progress sets. For four different population sizes  \gls{amount-of-seen-nodes}, we gradually increase the number of malicious nodes \(\gls{desired-malicious-node-tolerance}\) to  compare \(\gls{dss}\) with \(\gls{pss}\), and \(\gls{dps}\) with \(\gls{pps}\). The results presented in~\cref{fig:deterministic-vs-probabilistic-set-sizes-with-the-increase-of-malicious-nodes} lead us to the following conclusions:

\begin{itemize}
  \item The sizes of probabilistic sets \gls{pss} and \gls{pps} are much smaller compared to the sizes of  \gls{dss} and \gls{dps};
  \item The size of deterministic sets is, as expected, linearly dependant on the number of malicious network nodes, while the size of probabilistic sets shows sublinear growth. For certain parameters (e.g., when \gls{desired-malicious-node-tolerance} is reasonably small), the probabilistic set sizes are significantly smaller compared to their deterministic counterparts. The observed difference is significant for larger populations, when a probabilistic set size can be several orders of magnitude smaller compared to its deterministic counterpart.
\end{itemize}

Thus, we can conclude that probabilistic honest sets bring significant benefits for solving the problem of identifying honest nodes within subsets of network nodes, especially for large networks.

\section{Probabilistic set size}\label{sec:probabilistic-set-size}

% Intro.
Since node \gls{new-node} uses members of \gls{psh} (i.e., \gls{pss} or \gls{pps}) for transient or persistent communication, reducing \(|\gls{psh}|\) will diminish the communication complexity of our algorithm. As shown in the previous section, the size of \gls{psh} can be reduced by increasing the number of successes in the population. In this section, we examine the rate at which \(|\gls{psh}|\) is reduced with respect to the growth of \gls{amount-of-seen-nodes}, and compare \(|\gls{psh}|\) to sublinear functions of \gls{desired-malicious-node-tolerance} in the context of the current size of the Bitcoin IPV4 network.

% Our methodology.
An experiment was conducted to mimic algorithm execution that creates an increasing set of discoverable network nodes: We varied the number of discoverable nodes \gls{amount-of-seen-nodes} from \(2\) to \(20480\), which is the maximum number of total nodes that a Bitcoin client can store~\cite{biryukov2014deanonymisation}. For each \gls{amount-of-seen-nodes}, \gls{desired-malicious-node-tolerance} was set to \(\gls{amount-of-seen-nodes}-1\) and gradually decreased in increments of \(1\). During each experimental run,  the ratio between \gls{amount-of-seen-nodes} and \gls{desired-malicious-node-tolerance} was marked  when the following predicates (i.e., conditions) were met:

%Ovo su zahtjevi na veličinu vjerojatnosnog skupa u odnosu na pretpostavljeni broj malicioznih čvorova. Želimo bolje od k+1 tj.(2k+1), pa su ovo funkcije koje su sublinerane te bi nam dale puno bolje rezultate od linearnih.
\begin{multicols}{2}
  \begin{enumerate}
    \item \(|\gls{pss}| \leq  \sqrt{\gls{desired-malicious-node-tolerance}}\).
    \item \(|\gls{pss}| \leq \ln{\gls{desired-malicious-node-tolerance}}\).
    \item \(|\gls{pps}| \leq \sqrt{\gls{desired-malicious-node-tolerance}}\).
    \item \(|\gls{pps}| \leq \ln{\gls{desired-malicious-node-tolerance}}\).
  \end{enumerate}
\end{multicols}
% The results of the experiment
\begin{figure}[ht!]
  \centering
  \includegraphics[width=\linewidth]{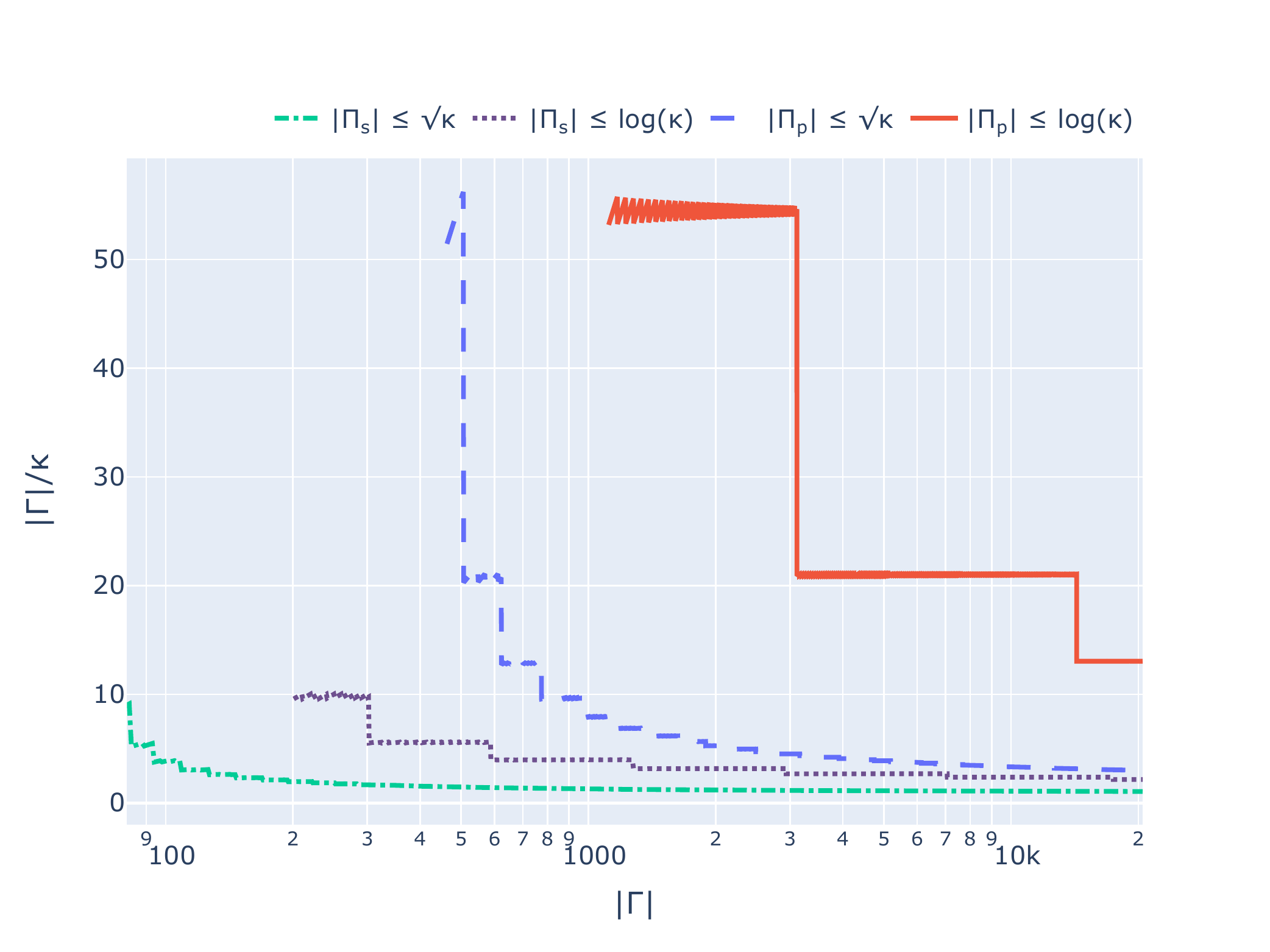}
  \caption{Ratio \(\frac{\gls{amount-of-seen-nodes}}{\gls{desired-malicious-node-tolerance}}\) when the four predicates limiting the probabilistic set sizes to \(  \sqrt{\gls{desired-malicious-node-tolerance}}\) and \( \ln{\gls{desired-malicious-node-tolerance}}\) have been satisfied with an increase of \gls{amount-of-seen-nodes}.}\label{fig:ratio-total-nodes-vs-malicious-nodes-when-complexity-converges-with-increase-of-nodes}
\end{figure}

% Interpretation of results.
Results are depicted in~\cref{fig:ratio-total-nodes-vs-malicious-nodes-when-complexity-converges-with-increase-of-nodes}. By examining the ratios indicating potential reduction of probabilistic set sizes with respect to malicious node tolerance, we can conclude that as the size of the population increases, the ratio between \gls{amount-of-seen-nodes} and \gls{desired-malicious-node-tolerance} required to reduce the probabilistic set sizes decreases. Thus, it is easier to satisfy the four predicates limiting \(|\gls{psh}|\) for larger populations, i.e., for larger sets of discoverable nodes.

% Context of results in a real-world use case of Bitcoin.
\cref{tab:convergence-and-utility-data-for-the-network-size-of-6356} shows the values extracted from~\cref{fig:ratio-total-nodes-vs-malicious-nodes-when-complexity-converges-with-increase-of-nodes} for \(\gls{amount-of-seen-nodes}=6356\), which is the number of active Bitcoin IPV4 nodes discovered and reported in~\cite{deshpande2018btcmap}. This number is used as the default network size in all the following experiments.

% https://www.tablesgenerator.com/#
\begin{table}[ht!]
  \caption{\label{tab:convergence-and-utility-data-for-the-network-size-of-6356}\(ps\): probabilistic set type, \(f\): upper bound for probabilistic set size as a function of \gls{desired-malicious-node-tolerance}, \(f(\gls{desired-malicious-node-tolerance})\): \(f\) evaluated at respective \gls{desired-malicious-node-tolerance}, \(P\): predicate, \(|\gls{ps}|\): respective probabilistic set size, \(|\gls{ds}|\): respective deterministic set size}
  \begin{tabular}{@{}|c|c|c|c|c|@{}}
    \toprule
    \(\gls{amount-of-seen-nodes}\)                                        & \multicolumn{4}{c|}{\(6356\)}                                                                                                                                                                                                         \\ \midrule
    \(ps\)                                                                & \multicolumn{2}{c|}{\(\gls{pss}\)}                      & \multicolumn{2}{c|}{\(\gls{pps}\)}                                                                                                                                          \\ \midrule
    \(f\)                                                                 & \(sqrt(\gls{desired-malicious-node-tolerance})\)        & \(\ln(\gls{desired-malicious-node-tolerance})\)         & \(sqrt(\gls{desired-malicious-node-tolerance})\)        & \(\ln(\gls{desired-malicious-node-tolerance})\)         \\ \midrule
    \(P\)                                                                 & \(|ps| \leq f(\gls{desired-malicious-node-tolerance})\) & \(|ps| \leq f(\gls{desired-malicious-node-tolerance})\) & \(|ps| \leq f(\gls{desired-malicious-node-tolerance})\) & \(|ps| \leq f(\gls{desired-malicious-node-tolerance})\) \\ \midrule
    \(\gls{amount-of-seen-nodes}-\gls{desired-malicious-node-tolerance}\) & 549                                                     & 3985                                                    & 4615                                                    & 6053                                                    \\ \midrule
    \gls{desired-malicious-node-tolerance}                                & 5807                                                    & 2371                                                    & 1741                                                    & 303                                                     \\ \midrule
    \(\gls{amount-of-seen-nodes}/\gls{desired-malicious-node-tolerance}\) & 1.09454                                                 & 2.68073                                                 & 3.65078                                                 & 20.9769                                                 \\ \midrule
    \(f(\gls{desired-malicious-node-tolerance})\)                         & 76.20367                                                & 7.77107                                                 & 41.72529                                                & 5.71373                                                 \\ \midrule
    \(|\gls{ds}|\)                                                        & 5808                                                    & 2372                                                    & 3483                                                    & 607                                                     \\ \midrule
    \(|\gls{ps}|\)                                                        & 76                                                      & 7                                                       & 41                                                      & 5                                                       \\ \midrule
    \(\frac{|\gls{ds}|}{|\gls{ps}|}\)                                     & \textbf{76.421052}                                      & \textbf{338.85714}                                      & \textbf{84.951219}                                      & \textbf{121.4}                                          \\ \midrule
    \(p\)                                                                 & 0.9990005                                               & 0.9990004                                               & 0.9990073                                               & 0.9990014                                               \\ \bottomrule
  \end{tabular}

\end{table}

% Example of how to read the table.
For example, by examining the first column in \cref{tab:convergence-and-utility-data-for-the-network-size-of-6356}, we see that to satisfy the predicate \(|\gls{pss}| \leq sqrt(\gls{desired-malicious-node-tolerance})\), even up to \(5807\)  of \(6356\) network nodes can be malicious, while the size of \gls{pss} is less than or equal to \(\sqrt{5807}=76.20367\), which corresponds to \(76\), and the correct execution is guaranteed with \(p=0.9990005\). If we would want to deterministically find at least one honest node within the same network, \(|\gls{dss}|= \gls{desired-malicious-node-tolerance}+1 = 5808\). In other words, \(76.421052\) times more nodes need to be queried for a deterministic answer in comparison to a probabilistic answer.

% Constraint on \gls{amount-of-seen-nodes}.
From a pragmatic point of view, the first three predicates limiting probabilistic set sizes with respect to \gls{desired-malicious-node-tolerance} (results displayed in the first three columns of~\cref{tab:convergence-and-utility-data-for-the-network-size-of-6356}) can be satisfied relatively easily, while the last condition has little value for a real-world scenario due to its extremely high \(\gls{amount-of-seen-nodes}/\gls{desired-malicious-node-tolerance}\) ratio. Further reduction of the probabilistic set size requires an even larger \(\gls{amount-of-seen-nodes}/\gls{desired-malicious-node-tolerance}\) ratio and is considered impractical. Therefore, we decided to use the square root as a sublinear function that bounds the probabilistic set size with respect to malicious node tolerance and we define the following constraint which will be of great use in the upcoming complexity analysis:

\begin{constraint}\label{con:sqrt-psi-max-size}
  \(|\gls{psh}|\) is bounded by \(\sqrt{\gls{desired-malicious-node-tolerance}}\), i.e., \(|\gls{psh}| \leq \sqrt{\gls{desired-malicious-node-tolerance}}\).
\end{constraint}

% Outro.
Note that so far we have relied on the existence of the set \gls{unique-nodes-found-in-gathering} without elaborating on how this set is constructed, as our analysis has been independent of any particular DLT solution or underlying network topology. However, both the topology and specifics of DLT solutions need to be considered when defining the Aurora algorithm, as we explain in the following section.

\section{The Aurora algorithm and its applications}\label{sec:collecting-a-subset-of-the-network}

The Aurora algorithm is used to construct probabilistic sets from a list of discoverable nodes. When performing the algorithm, node \gls{new-node} contacts other network nodes: In accordance with~\cref{con:no-bootstrap-peers} it does not have access to trusted nodes and contacts a random one. As previously stated, starting from node \gls{first-contact}, node \gls{new-node} explores the network in a series of steps, where a single step is defined as:

% What is a \gls{draw}?
\begin{definition}[\Gls{draw}]\label{def:draw}
  A \gls{draw} is a step performed by node \gls{new-node} during network exploration which consists of two messages: a ping message from node \gls{new-node} to a network node requesting its peer list, and the corresponding pong message containing a response denoted as \gls{peer-list-response}. A \gls{draw} at step \idx{0} expands \(\gls{unique-nodes-found-in-gathering}_{\idx{0}-1}\) with the unique nodes found in \(\gls{peer-list-response}_{\idx{0}}\), i.e., \(\gls{unique-nodes-found-in-gathering}_{\idx{0}}\gets \gls{unique-nodes-found-in-gathering}_{\idx{0}-1} \cup \gls{peer-list-response}_{\idx{0}}\)
\end{definition}

A network node can receive multiple ping queries and is assumed to be stateful~\cite{miller2015discovering}, which means it retains state about which addresses of neighboring peers are revealed to node \gls{new-node} together with a timestamp. Consequently, the network node will not reveal nodes that have already been revealed to node \gls{new-node} in a previous pong message. A network node is no longer queried by node \gls{new-node} if it responds with an empty peer list or fails to respond. Once a \gls{draw} is successful, node \gls{new-node} terminates a connection to the network node since resources should be released as soon as possible. \Glspl{draw} are executed in steps until a \textit{halting condition} is satisfied.

% What is a halting condition?
\begin{definition}
  [Halting condition]\label{def:halting-condition}

  Halting condition is a predicate that defines specific conditions which indicate to node \gls{new-node} to terminate further \glspl{draw} in the network.
\end{definition}

Node \gls{new-node} checks, after each \gls{draw}, whether the halting condition is satisfied and consequently terminates further \glspl{draw}. Hereafter, unless otherwise specified, the \textit{default halting condition} is defined as a step when all discoverable nodes stop responding to ping requests of node \gls{new-node} (i.e., there are no available nodes left to query).

We define a sequence of \gls{draw} steps as follows:
% What is a \gls{gathering}?
\begin{definition}[\Gls{gathering}]\label{def:gathering}

  \Gls{gathering} is the process of network exploration which consists of a sequence of \glspl{draw} executed in steps \(\idx{0}=1\ldots\gls{number-of-draws}\), where a halting condition is met in step \gls{number-of-draws}. We assume it is executed in a \textbf{directed} and \textbf{acyclic} manner.
\end{definition}
We assume that a network can be represented as a graph in which nodes are represented as vertices and connections between nodes are represented as edges, where edges have an associated direction. In such a graph, a \gls{gathering} can be associated with a Directed Acyclic Graph (DAG) consisting of nodes contacted during a \gls{gathering}. After a node responds with an empty peer list or does not respond at all, a subsequent node is selected from \gls{unique-nodes-found-in-gathering} uniformly at random.
As node \gls{new-node} performs the \gls{gathering}, it also overlooks the creation of the associated DAG to ensure that cycle formation is avoided. %since the Aurora algorithm is executed only on the new node \gls{new-node}, which means that \gls{new-node} has complete knowledge of the generated directed graph and complete control over the candidate for the next \gls{draw}.

% Comparison of a \gls{gathering} with a scraper.
Although at first glance a \gls{gathering} is similar to the process of recursively scraping the entire network, it is something quite different. The purpose of a \gls{gathering} is to collect a minimal number of nodes within a minimal period of time to construct probabilistic honest sets.
The probabilistic sets are then used by node \gls{new-node} to choose an optimal bootstrap candidate and honest nodes for transaction inclusion verification.

The Aurora algorithm, as defined in~\cref{alg:aurora-pseudocode}, requires the following input parameters: a first contact node \gls{first-contact}, a desired malicious node tolerance \gls{desired-malicious-node-tolerance}, the probability \gls{correctness-probability} under which it guarantees correct execution, a halting condition \gls{halting-condition-encapsulation}, the maximum acceptable size of the probabilistic set \(|\gls{psh}|\), and the number of honest nodes \idx{2} which the probabilistic honest set should contain. The last parameter consequently determines which of the probabilistic sets is created: \gls{pss} or \gls{pps}. The algorithm performs a do-while loop which either constructs a \gls{psh} (line \(7\)) or throws an exception (line \(13\)). Ping and pong messages are exchanged between node \gls{new-node} and a randomly sampled responding node from \gls{unique-nodes-found-in-gathering} (lines \(9-11\)). Unique nodes gathered from ping responses are used to expand \gls{unique-nodes-found-in-gathering} (line \(12\)). After each ping and pong exchange, the algorithm checks weather the halting condition has been satisfied in line \(13\).

\begin{algorithm}[ht!]
  \Input{
    \gls{correctness-probability} --- Probability guarantee
  }

  \Input{
    \gls{first-contact} --- First contact node identifier
  }

  \Input{
    \gls{halting-condition-encapsulation} --- Halting condition encapsulation
  }
  \Input{
    \(|\gls{psh}|\) --- Maximum \gls{psh} size
  }
  \Input{
    \idx{2} --- \([1, \lfloor |\gls{psh}| /2  \rfloor + 1]\)
  }
  \Input{
    \gls{desired-malicious-node-tolerance} --- Desired malicious node tolerance
  }

  \(\gls{next-draw-in-gathering} \gets \gls{first-contact}\)\;
  \(\gls{unique-nodes-found-in-gathering} \gets \emptyset\)\;

  \Do{\(True\)}{
    \(
    \gls{generic-random-variable} \sim Hypergeometric(
    |\gls{unique-nodes-found-in-gathering}|,
    |\gls{unique-nodes-found-in-gathering}|-\gls{desired-malicious-node-tolerance},
    |\gls{psh}|
    )
    \)\;

    \uIf{\(p(X \geq \idx{2}) \geq \gls{correctness-probability} \)}{
      \(\gls{psh} \gets\) Sample without replacement \(|\gls{psh}|\) nodes from \gls{unique-nodes-found-in-gathering}\;
      \textbf{return} \gls{psh}
    }
    \Else{
      \(\gls{next-draw-in-gathering} \gets \text{random responding node in}~\gls{unique-nodes-found-in-gathering}\)\;
      Send ping to \gls{next-draw-in-gathering}\;
      \(\gls{peer-list-response} \gets \text{peers in pong message from}~\gls{next-draw-in-gathering}\)\;
      Add \gls{peer-list-response} to \gls{unique-nodes-found-in-gathering}\;

      \If{\(\gls{halting-condition-encapsulation}.isHaltingConditionReached(\gls{next-draw-in-gathering}, \gls{peer-list-response})\)}{
        \textbf{throw}
      }
    }
  }
  \caption{Aurora algorithm}\label{alg:aurora-pseudocode}
\end{algorithm}

% The halting condition API
The halting condition used in line \(12\) is an injected dependency that implements a generic method which returns \(true\) if a \gls{gathering} should halt, or \(false\) otherwise. It uses as input parameters \gls{next-draw-in-gathering}, which is the next node queried for its peer list, and the corresponding peer list \gls{peer-list-response}.

%\begin{enumerate}
%  \item \(isHaltingConditionReached(\gls{next-draw-in-gathering}, \gls{peer-list-response})\)
%        \begin{itemize}
%          \item Params
%                \begin{enumerate}
%                  \item \gls{next-draw-in-gathering} --- responder identifier.
%                  \item \gls{peer-list-response} --- response from \gls{next-draw-in-gathering}.
%                \end{enumerate}
%          \item Returns
%                \begin{itemize}
%                  \item \(true\) if \gls{gathering} should halt, \(false\) otherwise.
%                \end{itemize}
%        \end{itemize}
%\end{enumerate}

% Halting condition examples.
A halting condition may be defined as follows:
\begin{enumerate}
  \item halt if the average number of newly-discovered nodes in a \gls{draw} falls below a defined threshold, or
  \item halt if a predefined maximal number of \gls{gathering} steps has been reached.
\end{enumerate}

Examples of halting conditions and their influence on algorithm performance are investigated in~\cref{sec:methodology-and-experiments}.

\subsection{Choosing a bootstrap candidate}\label{sec:algorithm-pseudocode}

%Ethereum P2P Networking / Sharding by Felix Lange and Péter Szilágyi in Taipei, March 2018
% https://www.youtube.com/watch?v=qJA6J0mP73w&t=575s
% Why is pss chosen for bootstrap peer detection
For the purposed of identifying an honest bootstrap candidate, node \gls{new-node} should use the probabilistic safe set generated as the output of the Aurora algorithm since it contains at least \(1\) honest node with high probability. Here, we assume that nodes in the network advertise the state of the ledger (e.g., the highest total difficulty represents the latest ledger in Ethereum), and node \gls{new-node} will attempt to synchronize its ledger with other nodes using the order starting from the latest to the oldest advertised state. To understand why \gls{pss} is adequate for this purpose, let us analyze the behavior of an adversary. First, if the adversary advertises an older ledger state compared to honest nodes (i.e., a lower chain head), node \gls{new-node} will synchronize with another honest member of \gls{pss} that reveals the latest ledger state (i.e., the highest chain head). Second, if the adversary advertises a newer ledger state compared to an honest node (i.e., a highest chain head), then by~\cref{con:honest-majority} the ledger offered by the adversary must be tampered with. In this scenario, \gls{pss} will eventually provide safety with respect to blockchain synchronization. In other words, node \gls{new-node} could choose a malicious node from \gls{pss} and start synchronizing, but it will eventually detect that fraudulent data is provided by the malicious node by examining the ledger. Since there is a guarantee with probability \gls{correctness-probability} that an honest node is available among all nodes in \gls{pss}, the node will start synchronizing with the available honest node.

% Bootstrap peer detection pseudocode.
\cref{alg:bootstrap-peer-detection-pseudocode} relies on the existence of a \gls{pss} constructed by~\cref{alg:aurora-pseudocode} and allows node \gls{new-node} to choose a candidate node for synchronization. The candidate is the node that provides the most recent ledger according to the consensus protocol, e.g., advertises the longest chain, highest difficulty, etc. For each such candidate, node \gls{new-node} attempts to synchronize with its ledger. If it turns out that the ledger has been tampered with, node \gls{new-node} chooses the next candidate. The algorithm stops when node \gls{new-node} downloads and verifies an entire ledger without detecting any tampering.

\begin{algorithm}[ht!]

  \Input{
    \gls{pss} --- Probabilistic Safe Set
  }

  \(sync \gets false\)\;% chktex 21
  \While()
  {
    \(sync = false\)
  }
  {
    \(candidate \gets null\)\;% chktex 21
    \For{\(node\) in \gls{pss}}{
      \(candidate \gets\)\ node offering latest ledger state % chktex 21
    }

    Attempt to sync with \(candidate\)\;% chktex 21
    \uIf{Attempt successful}{
      \(sync \gets true\)\;% chktex 21
    }\Else(){
      Remove \(candidate\) from \gls{pss}\;
    }
  }
  \caption{\(\gls{pss}\) use for synchronization}\label{alg:bootstrap-peer-detection-pseudocode}
\end{algorithm}

\subsection{Transaction inclusion verification}

% Assumption necessary for the use of \gls{pps}
For the purpose of transaction inclusion verification, node \gls{new-node} should use the probabilistic progress set \gls{pps} to verify whether a transaction with identifier \gls{transaction-id} is contained within a block with identifier \gls{block-id}. Here we make further assumptions:

% The existence of integrity validating structures
\begin{assumption}\label{con:existence-of-integrity-validating-structures}
  The data contained in a block is part of an integrity-validating structure, and a node can verify that the transaction is indeed contained in that structure. It is assumed that all necessary additional metadata is available to the node.\footnote{Such structures are present in prominent DLT solutions in the form of Merkle trees and Merkle proofs.}
\end{assumption}

% Why is pps chosen for transaction inclusion verification
To explain the usefulness of a progress set \gls{pps} for transaction inclusion verification, we compare it with \gls{pss}. When relying on an instance of \gls{pss}, node \gls{new-node} cannot assume that the chosen bootstrap candidate is honest, since ledgers need to be downloaded and verified. However, node \gls{new-node} can compare answers from all members of \gls{pps} and conclude that the answer provided by the majority is true with probability \gls{correctness-probability}.

% Explanation of the algorithm.
When~\cref{con:existence-of-integrity-validating-structures} holds,~\cref{alg:transaction-inclusion-verification-pseudocode} either checks whether a transaction with ID \gls{transaction-id} is included in a block \gls{block-id} with probability \gls{correctness-probability} by checking with every member of \gls{pps} if the provided Merkle proof is valid with respect to the available Merkle root \gls{merkle-root}, and thus chooses the majority answer.

\begin{algorithm}[ht!]
  \Input{
    \gls{merkle-root} --- String, merkle root
  }
  \Input{
    \gls{transaction-id} --- String, transaction ID
  }
  \Input{
    \gls{block-id} --- String, block ID
  }
  \Input{
    \gls{pps} --- Probabilistic Progress Set
  }

  \(majority \gets \lfloor |\gls{pps}| /2  \rfloor + 1\)\;% chktex 21
  \(included \gets 0\)\;% chktex 21
  \(notIncluded \gets 0\)\;% chktex 21
  \For{\(node\) in \gls{pps}}{

    Request Merkle proof from \(node\)\;
    Verify Merkle proof for \gls{transaction-id} in \gls{block-id} using \gls{merkle-root}\;

    \uIf{\(node\) offers valid Merkle proof}{
      \(included \gets included+1\)\; % chktex 21
    }\Else(){
      \(notIncluded \gets notIncluded+1\)\; % chktex 21
    }
    \uIf{\(included = majority\)}{
      Declare transaction included\;
      \Return\;% chktex 21
    }\ElseIf{\(notIncluded = majority\)}{
      Declare transaction not included\;
      \Return\;% chktex 21
    }
  }
  \caption{\gls{pps} use for \gls{transaction-id} inclusion verification}\label{alg:transaction-inclusion-verification-pseudocode}
\end{algorithm}

% Outro.
%We continue by explaining how the required variables for both algorithms are initialized or obtained, as well as highlighting some relevant use cases.

\subsection{Initialization of parameters and use cases}\label{sec:variable-initialization-and-use-cases}

% Bootstrap peer detection variable initialization.
To use~\cref{alg:aurora-pseudocode}, at least six parameters are required. As mentioned earlier, it is recommended to set the correctness probability \gls{correctness-probability} to \(0.999\), with a note that the impact of \gls{correctness-probability} on the size of probability sets is part of our future work. The first contact node \gls{first-contact} can be found in any way a user sees fit, e.g., via chat, forums, etc. We define the default halting condition \gls{halting-condition-encapsulation} as the moment when there are no more nodes to ask for peer lists, although we strongly suggest changing this to a condition that better fits the underlying network topology and the resources of a ledger client (an example is given in~\cref{exp:average-number-of-new-nodes-discovered-in-a-draw}). The maximum size of \gls{psh} can be set in accordance with~\cref{con:sqrt-psi-max-size}, i.e., \(\lfloor\sqrt{\gls{desired-malicious-node-tolerance}}\rfloor\). The number of honest nodes \idx{2} in \gls{psh}, i.e.\ choosing between \gls{pss} and \gls{pps}, depends on the use case, as discussed further in this section.

% Initialization of \gls{desired-malicious-node-tolerance} - guess above or below.
Initializing \gls{desired-malicious-node-tolerance} is nontrivial: if a user makes a correct guess and sets \(\gls{desired-malicious-node-tolerance} = \gls{number-of-malicious-nodes-in-the-network}\), the algorithm executes optimally. If the user sets \(\gls{desired-malicious-node-tolerance} <\gls{number-of-malicious-nodes-in-the-network}\), then correct algorithm execution cannot be guaranteed, while if \(\gls{desired-malicious-node-tolerance} >\gls{number-of-malicious-nodes-in-the-network}\), correct execution is guaranteed with probability \gls{correctness-probability} at the cost of higher communication complexity. Thus, this parameter should in general be an overestimation of the envisioned number of malicious nodes in the network.

% Time-based API as a friendly facade.
Since it is excessive to expect that a number of malicious nodes in the network can be guessed correctly, an alternative can be offered instead of \gls{desired-malicious-node-tolerance} which uses another parameter, the maximum execution time in seconds \gls{maximum-execution-time} specifying how long a user is willing to wait for the algorithm to complete.
% How does the time-based API work?
Within a period of \gls{maximum-execution-time}, a node running the algorithm collects nodes. After the specified timeout occurs, using the values from~\cref{fig:ratio-total-nodes-vs-malicious-nodes-when-complexity-converges-with-increase-of-nodes}, the user is presented with a table such as~\cref{tab:convergence-and-utility-data-for-the-network-size-of-6356} to choose an adequate probabilistic set based on the underlying use case. %Since the time-based API is just a transitive way of expressing the desired tolerance to malicious nodes, the use and implications of such an API remain part of a future work.

% Initialization of variables for transaction inclusion verification.
To use~\cref{alg:bootstrap-peer-detection-pseudocode}, a user must obtain the transaction ID \gls{transaction-id}, the block ID \gls{block-id}, and the corresponding Merkle root \gls{merkle-root} from an entity issuing the transaction. For example, in a use case scenario when a merchant ships products after a buyer has paid for the goods, the buyer sends the above parameters to the merchant as proof of payment.

% Relevant Use Cases - \gls{pss}
The Aurora algorithm has a variety of potential use cases. For example, if a user is willing to download the ledger and time is not critical\footnote{The user is willing to wait for an extended period of time to collect unique nodes during a \gls{gathering}. The constructed \gls{psh} does not need to satisfy~\cref{con:sqrt-psi-max-size}}. The vital parameter for this use case is tolerance to \gls{desired-malicious-node-tolerance} malicious nodes, and thus the user may decide to construct a \gls{pss} that guarantees at least one honest node while specifying an overestimated \gls{desired-malicious-node-tolerance}.

% Relevant Use Cases - \gls{pps}
Similarly, if consumer-grade hardware is used to check the state of a transaction, but it cannot store the entire ledger, time is again not critical and a user wants a trustless verification without relying on a third-party service. For this use case, the user will opt to use the \gls{pps} to query the state of a transaction.

% Relevant use cases - restricted \gls{pss}
Finally, if a user applies a resource-constrained device (e.g., a mobile phone) which cannot store the entire ledger and is unwilling to download the header chain, but the user still wants to verify that a transaction is contained in a ledger and is interested in doing so in an efficient manner, the user should construct a \gls{pps} that follows constraints adequate for this use case, for example~\cref{con:sqrt-psi-max-size}. Since in this paper we focus on applying the Aurora algorithm on resource-constrained devices, this use case is discussed in more detail in~\cref{sec:transaction-inclusion-verification-for-light-clients}.

% Outro.
%Now that we have a better understanding of the algorithm not only in the context of set creation, but also in a DLT environment, we move on to the methodology of algorithm verification. Verifying our algorithm in a DLT production network is dubious at best --- including malicious actors in a production network is resource intensive, detrimental, and mostly useless since we would need to know the absolute state of the network to interpret the results. Therefore, testing in a simulation is a natural choice. For such a simulation, a network topology had to be modelled.

\section{Communication complexity}\label{sec:communication-complexity}

Let us examine an upper bound on the number of messages exchanged by node \gls{new-node} running the Aurora algorithm as a function of the desired malicious nodes tolerance \gls{desired-malicious-node-tolerance}.

The communication complexity of the algorithm can be decomposed into two main parts:
\begin{enumerate}
  \item the number of messages exchanged during a \gls{gathering}, and
  \item the number of messages exchanged during communication with members of \gls{psh}.
\end{enumerate}

The total communication complexity can be expressed as a sum of these values, which we can derive from the following three equations.

% \Gls{gathering} complexity - average number of nodes per \gls{draw}.
First, a \gls{gathering} can be described as a linear function where the domain is the number of steps performed during a \gls{gathering} \gls{number-of-draws}, the co-domain is the number discoverable nodes discovered during the \gls{gathering} \gls{amount-of-seen-nodes}, and the slope \gls{unique-nodes-per-draw-random-variable} is a random variable. Thus, we write

\begin{equation}
  \label{eqn:gathering-as-a-function-of-draws}
  \gls{amount-of-seen-nodes} = f(\gls{number-of-draws}) =  \gls{unique-nodes-per-draw-random-variable} * \gls{number-of-draws}.
\end{equation}

% *** USEFUL BOOKMARKS ***
% We should calculate the mean of the means
% 7. Confidence Intervals-rUxP7TM8-wo @ 00:24:19
% Quote:

% If this set is sufficiently large --- certainly 1 is not sufficiently large --- then it will be the case that the mean of the means --- so I take the mean of each sample and then I can now plot all of those means and so take the mean of those, right --- and  they'll be normally distributed.

%https://www.investopedia.com/terms/c/central_limit_theorem.asp

% The complexity of the \gls{gathering} - mean of the slope averages.
The average slope \gls{average-unique-nodes-per-draw-random-variable} is a random variable  and is the average of the sample means\footnote{Central Limit Theorem (CLT)
} that can be calculated for any network or topology via Monte Carlo simulation to approximate a normal distribution. By applying the previous statement to~\cref{eqn:gathering-as-a-function-of-draws}, we obtain the following

\begin{equation}
  \label{eqn:gathering-as-a-function-of-draws-averaged}
  \gls{amount-of-seen-nodes} = f(\gls{number-of-draws}) =  \gls{average-unique-nodes-per-draw-random-variable} * \gls{number-of-draws}.
\end{equation}

% The complexity of the \gls{gathering} - total amount of nodes as a function of \gls{desired-malicious-node-tolerance}.
Second, as observed in~\cref{fig:ratio-total-nodes-vs-malicious-nodes-when-complexity-converges-with-increase-of-nodes},
the ratio \(\gls{ratio-ps-sub-linear-function-of-desired-malicious-node-tolerance} = \frac{\gls{amount-of-seen-nodes}}{\gls{desired-malicious-node-tolerance}}\) when \(|\gls{pss}|\) and \(|\gls{pps}|\) begin behaving like \( \sqrt{\gls{desired-malicious-node-tolerance}}\) is a function of \gls{desired-malicious-node-tolerance} and can be precomputed\footnote{E.g, for \gls{pps}, \gls{ratio-ps-sub-linear-function-of-desired-malicious-node-tolerance} ranges from \(55.22\) to \(1.99\) for all relevant values of the domain.}. Thus, the total unique number of nodes \gls{amount-of-seen-nodes} can also be expressed as:

\begin{equation}
  \label{eqn:unique-nodes-as-a-function-of-malicious-nodes}
  \gls{amount-of-seen-nodes} = f(\gls{desired-malicious-node-tolerance}) =
  \gls{ratio-ps-sub-linear-function-of-desired-malicious-node-tolerance} * \gls{desired-malicious-node-tolerance}
\end{equation}

By joining~\cref{eqn:gathering-as-a-function-of-draws-averaged} and~\cref{eqn:unique-nodes-as-a-function-of-malicious-nodes} we get

\begin{equation}
  %\gls{amount-of-seen-nodes} = f(\gls{number-of-draws})  = \gls{average-unique-nodes-per-draw-random-variable} * \gls{number-of-draws} \implies 
  \gls{ratio-ps-sub-linear-function-of-desired-malicious-node-tolerance} * \gls{desired-malicious-node-tolerance}  = \gls{average-unique-nodes-per-draw-random-variable} * \gls{number-of-draws} \implies \gls{number-of-draws} = \frac{\gls{ratio-ps-sub-linear-function-of-desired-malicious-node-tolerance}}{\gls{average-unique-nodes-per-draw-random-variable}} * \gls{desired-malicious-node-tolerance}
\end{equation}

Since there are \gls{number-of-draws} steps in a \gls{gathering} and each step involves ping and pong messages, the number of messages exchanged during a \gls{gathering} can be expressed as

\begin{equation}
  \label{eqn:complexity-of-the-gathering}
  f(\gls{desired-malicious-node-tolerance}, \gls{average-unique-nodes-per-draw-random-variable}) =  2 * \ceil[\bigg]{\frac{\gls{ratio-ps-sub-linear-function-of-desired-malicious-node-tolerance} * \gls{desired-malicious-node-tolerance} }{\gls{average-unique-nodes-per-draw-random-variable}}}
\end{equation}

% The complexity of the decision - maximum size of \gls{psh}.
Third, given~\cref{con:sqrt-psi-max-size}, the number of requests (pings) to be sent from the node executing the algorithm to the members of the constructed \gls{psh} can be expressed as follows:

\begin{equation}
  \label{eqn:pps-as-a-function-of-malicious-nodes}
  f(\gls{desired-malicious-node-tolerance}) = |\gls{psh}| = \lfloor\sqrt{\gls{desired-malicious-node-tolerance}}\rfloor
\end{equation}

In other words, the maximum number of messages exchanged when communicating with members of \gls{psh} is exchanged when all members \gls{psh} must be queried. In total, \(|\gls{psh}|\) ping and \(|\gls{psh}|\) pongs messages are exchanged, which can be expressed as follows:

\begin{equation}
  \label{eqn:complexity-of-the-decision}
  f(\gls{desired-malicious-node-tolerance}) =  2 * \lfloor\sqrt{\gls{desired-malicious-node-tolerance}}\rfloor
\end{equation}

% Summing up the two components.
Consequently, the total number of messages exchanged before the algorithm terminates can be expressed as the sum of~\cref{eqn:complexity-of-the-gathering} and~\cref{eqn:complexity-of-the-decision}:

\begin{equation}
  \label{eqn:total-number-of-exchanged-messages-before-a-decision-can-be-made}
  f(\gls{desired-malicious-node-tolerance}, \gls{average-unique-nodes-per-draw-random-variable}) =  2 * \ceil[\bigg]{\frac{\gls{ratio-ps-sub-linear-function-of-desired-malicious-node-tolerance} * \gls{desired-malicious-node-tolerance} }{\gls{average-unique-nodes-per-draw-random-variable}}} + 2 * \lfloor\sqrt{\gls{desired-malicious-node-tolerance}}\rfloor
\end{equation}

where \gls{ratio-ps-sub-linear-function-of-desired-malicious-node-tolerance} can be read from \cref{fig:ratio-total-nodes-vs-malicious-nodes-when-complexity-converges-with-increase-of-nodes} and \gls{average-unique-nodes-per-draw-random-variable} can be set equal to the threshold on the average number of newly discovered nodes per \gls{draw} (see~\cref{exp:average-number-of-new-nodes-discovered-in-a-draw}).

\section{Methodology and experiments}\label{sec:methodology-and-experiments}
% *** USEFULL BOOKMARKS ***

% https://stackoverflow.com/questions/2140787/select-k-random-elements-from-a-list-whose-elements-have-weights?noredirect=1
Having established a theoretical basis for the Aurora algorithm performance, we verify our findings using simulations. In this section, we explain the applied methodology and present the results of five distinct experiments performed using the available open-source tools.

% Reference to previous work confirming valid network topology,
The network topology was modelled in accordance with previous work~\cite{mivsic2019modeling,deshpande2018btcmap}, i.e., by modelling a highly connected core network and a lightly connected edge. The active Bitcoin IPV4 network slice defined in~\cite{deshpande2018btcmap} was chosen as the basis for the experiments, i.e., all experiments use a Bitcoin-like network with a total of \(6356\) nodes.

% The Adversary.
An adversary was modelled to advertise a spoofed total difficulty in the network that is higher than the canonical chain difficulty. When asked about transaction inclusion, the adversary always lies. When an adversary replies with a peer list, it reveals only malicious peers. All malicious nodes know about each other (the assumption that a malicious clique is fully connected is supported by~\cite{dagon2007taxonomy}). Finally, a single group of colluding malicious peers was modelled in the network.

% The simulator of choice.
The Simblock~\cite{aoki2019simblock} simulator was applied to define a network topology and extended to enable nodes to send ping messages and respond with pong messages, as defined by the Bitcoin protocol~\cite{biryukov2014deanonymisation,deshpande2018btcmap,miller2015discovering,heilman2015eclipse,mivsic2019modeling}, where a ping message maps to a GETADDR message and a pong request message to an ADDR message. The simulator was redesigned to rely on the Java streaming API to reduce its memory requirements and allow parallel execution. All other parameters are inherited from the Simblock codebase. Mining did not occur during a simulation run (i.e., the chain header was not modified). The Algorithms~\ref{alg:aurora-pseudocode},~\ref{alg:bootstrap-peer-detection-pseudocode} and~\ref{alg:transaction-inclusion-verification-pseudocode} were redesigned to match the underlying discrete event-driven environment.

\subsection{Experiment 1: Algorithm effectiveness when increasing the percentage of malicious network nodes}\label{exp:efficiency}

% The aim of the experiment.
The goal of this experiment is to evaluate the effectiveness of the algorithm when gradually increasing the percentage of malicious network nodes, while the node running the algorithm is assumed to have unlimited time and resources. No upper bound is placed on the probabilistic set sizes (i.e., \cref{con:sqrt-psi-max-size} has been abolished).

% Execution Methodology.
The effectiveness of the algorithm is measured by running a Monte Carlo simulation of the network while gradually increasing the number of malicious nodes from \(0\%\) to \(100\%\). Malicious nodes are selected  uniformly at random from the existing nodes in the topology and marked as malicious. First contact nodes \gls{first-contact} are also selected uniformly and randomly from malicious nodes to ensure a \(99\%\) confidence level and a \(1\%\) margin of error. We use the default halting condition for each \gls{gathering}, i.e., the node executing the algorithm halts the \gls{gathering} if there are no more nodes to ask for peer lists. The ultimate goal of the node executing the algorithm is to synchronize with an honest peer. The desired malicious node tolerance  \gls{desired-malicious-node-tolerance} is always set equal to the maximum value \gls{number-of-malicious-nodes-in-the-network}.

% Outcome - Halt and progress.
An outcome of a simulation run is classified as \textbf{halt} if the algorithm cannot guarantee protection against a desired number of malicious nodes and the node decides to abort operation.
In contrast, an outcome is classified as \textbf{progressed} when the algorithm does not \textbf{halt}, i.e., it has succeeded in making a decision about a ledger synchronization candidate, which may be either honest or malicious.

% Outcome - Success.
An outcome is marked as \textbf{success} if the node has chosen to \textbf{halt} or has \textbf{progressed} to synchronize with an honest node. Otherwise it is marked as \textbf{failure} since the node executing the algorithm decided to synchronize with a malicious node.

\begin{figure}[ht!]
  \centering
  \includegraphics[width=\linewidth]{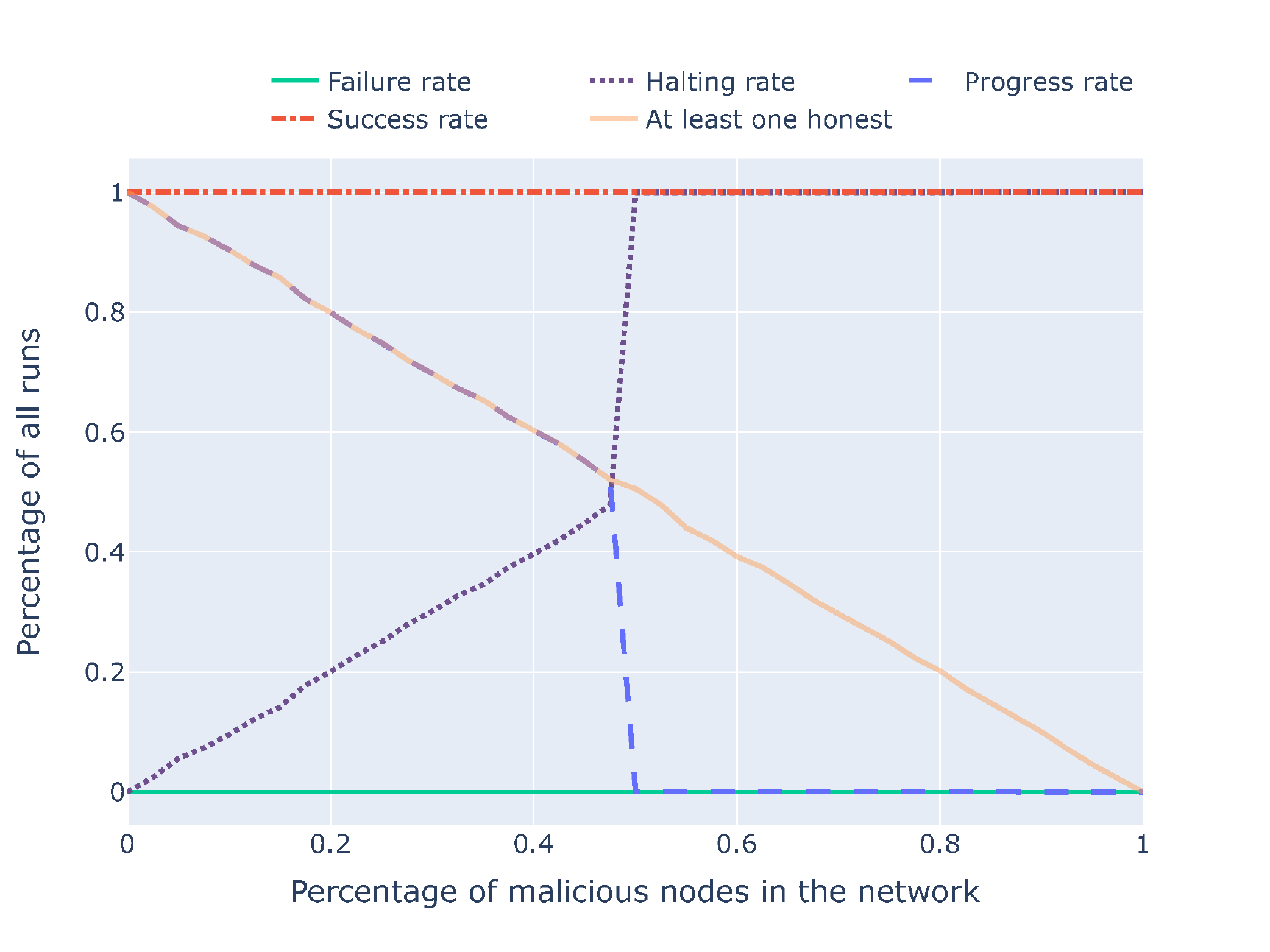}%
  \caption{Outcomes of the algorithm with an increase of the percentage of malicious network nodes.}\label{fig:outcomes-with-the-increase-of-malicious-nodes-in-the-network}
\end{figure}

% Interpretation of results.
The results are displayed in~\cref{fig:outcomes-with-the-increase-of-malicious-nodes-in-the-network}. The algorithm behaves as expected, i.e., it maintains a high success rate regardless of the underlying network topology or entry point, and identifies an honest node for persistent communication in \(99.9\%\) of cases. The algorithm consistently halts when the number of malicious nodes in the network exceeds \(50\%\) in accordance with~\cref{con:honest-majority}: It halts only when the first contact is malicious and the \gls{gathering} could not collect a desired number of nodes. %When the number of malicious nodes in the network exceeds \(50\%\), there is no longer a deterministic majority, i.e., no progress can be made, and the node halts consistently.

% Eventual safety.
The experiment highlights the idea of eventual safety with respect to ledger synchronization. As mentioned earlier, the algorithm provides eventual safety with respect to blockchain synchronization as long as the first contact is not malicious. That is, if the first contact is honest, the algorithm ensures the presence of at least one honest node (see trace \textit{At least one honest} in~\cref{fig:outcomes-with-the-increase-of-malicious-nodes-in-the-network}) and node \gls{new-node} will eventually detect that fraudulent data is provided by malicious nodes and start synchronization with the available honest node.

\subsection{Experiment 2: Number of discoverable nodes as a function of the number of draws}

% The aim of the experiment.
This experiment investigates how the number of \gls{gathering} steps \gls{number-of-draws} relates to the number of discoverable nodes, given an underlying network topology and without the presence of malicious nodes.

% Execution Methodology.
A total of \(1000\) experiments were conducted where node \gls{new-node} enters a network using a randomly-selected first contact node \gls{first-contact}. Node \gls{new-node} has unlimited time and resources, and the halting condition remains the default one.

\begin{figure}[ht!]
  \centering
  \includegraphics[width=\linewidth]{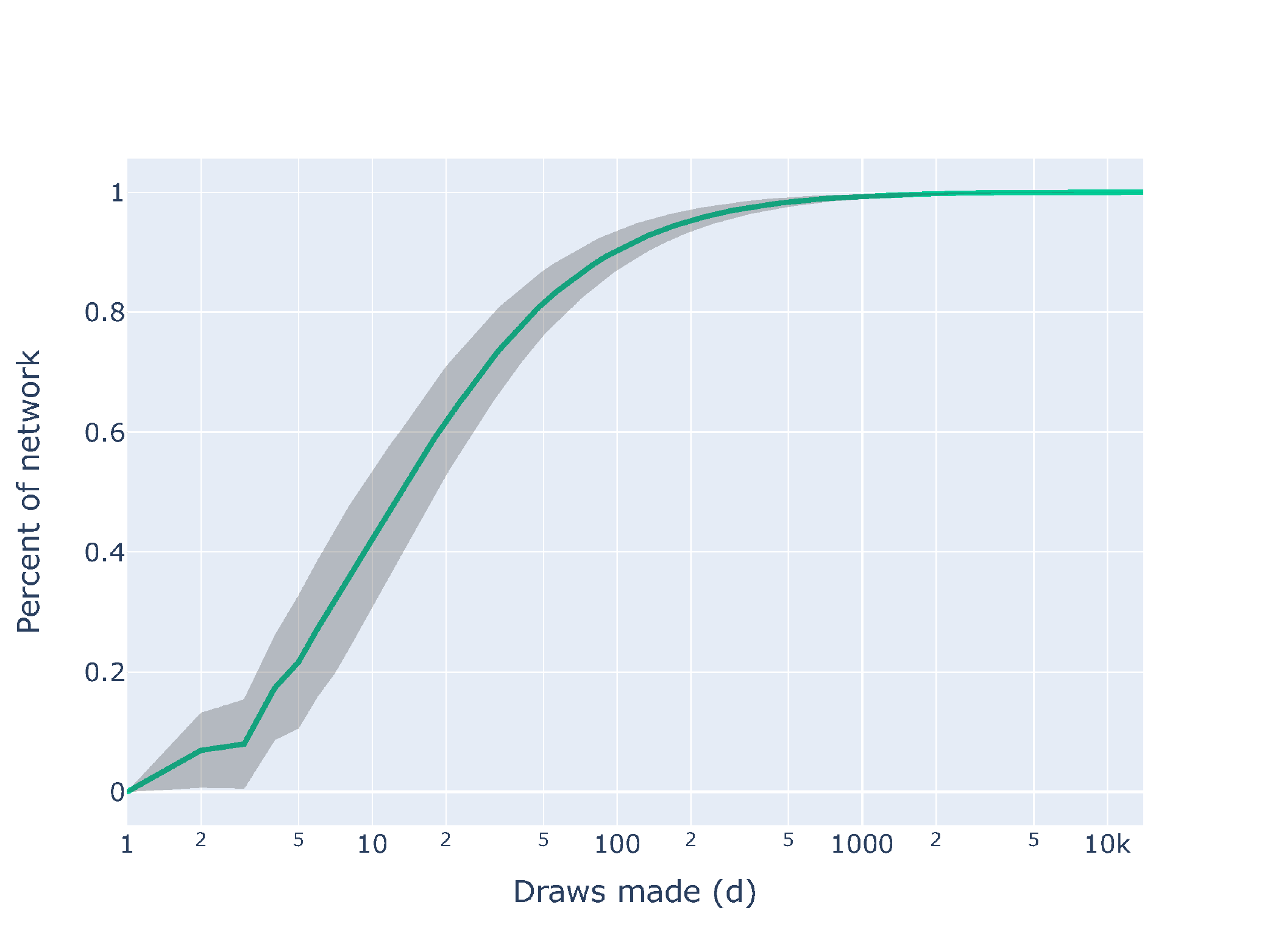}%
  \caption{The average and standard deviation of the node discovery rate with an increase of the number of \gls{gathering} steps during \(1000\) experiments.}\label{fig:average-percentage-of-seen-nodes-with-the-increase-of-draws}
\end{figure}

% Interpretation of results.
\cref{fig:average-percentage-of-seen-nodes-with-the-increase-of-draws} shows the average rate at which \gls{unique-nodes-found-in-gathering-in-step} grows as new \glspl{draw} are performed, i.e.\ the \textbf{node discovery rate} for the executed \(1000\) runs is depicted as a function of the number of \gls{gathering} steps. The shaded region around the average represents uncertainty and is bounded by one standard deviation. The results show that the node discovery rate is irregular and depends on the underlying topology and entry point. That is, if a node encounters a high-degree node early, the network expansion is faster and vice versa.

% Conclusions drawn from the results.
The results lead us to two conclusions. First, the number of newly discovered nodes at the beginning of a \gls{gathering} shows a linear dependence on the number of \gls{gathering} steps. As the \gls{gathering} progresses, the expansion rate slows down, which means that new nodes are harder to find. Second, after a certain point in time, the continuation of a \gls{gathering} makes little sense, as very few, if any, new nodes are discovered. From a pragmatic point of view, it is unreasonable to discover the entire network before the algorithm terminates. The communication complexity of the \gls{gathering} can be reduced by introducing a more complex halting condition which we investigate in the following experiment.

\subsection{Experiment 3: Average number of new nodes discovered in a \gls{draw}}\label{exp:average-number-of-new-nodes-discovered-in-a-draw}
% The aim of the experiment.
After studying the algorithm effectiveness when node \gls{new-node} has unlimited resources, we introduce additional constraints to test the algorithm in a more realistic scenario. This experiment examines whether a custom halting condition based on  the average number of newly discovered nodes can be constructed to reduce the communication complexity of the \gls{gathering} in terms of the number of \gls{gathering} steps, without significantly affecting the number of nodes discovered during a \gls{gathering}.

% Execution Methodology.
Node \gls{new-node} enters a network using a randomly-selected first contact node \gls{first-contact} and monitors the average number of newly discovered nodes in a \gls{draw} (at least \(10\) \glspl{draw} are needed before an average is calculated). If the average number of newly discovered nodes in a \gls{draw} falls below a threshold, the node halts. The threshold varies from \(1\) to \(500\), and for each threshold, a total of \(1000\) experiments is performed, based on which the average and standard deviation of the percentage of the network discovered is calculated.

\begin{figure}[ht!]
  \centering
  \includegraphics[width=\linewidth]{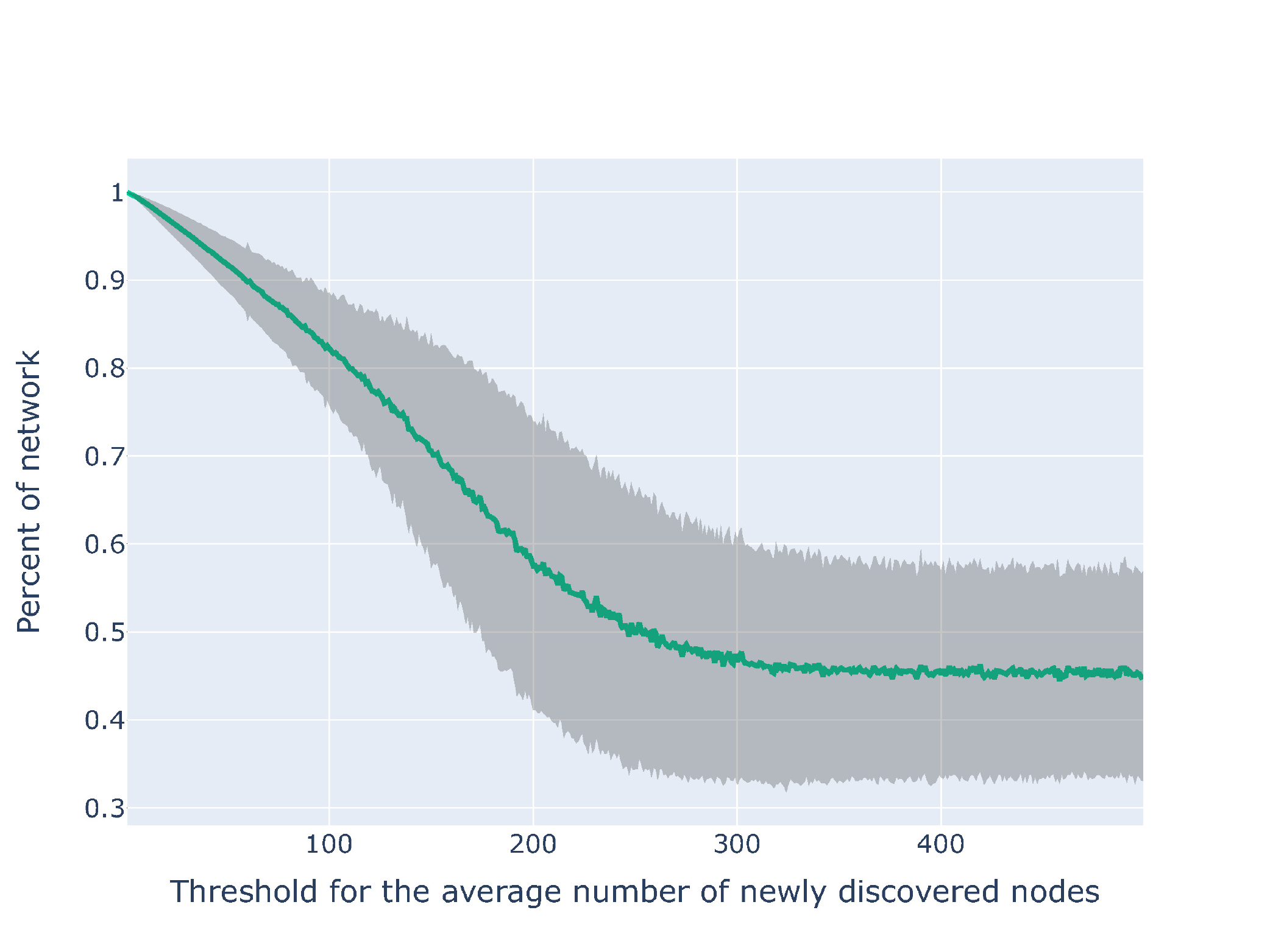}%
  \caption{The average and the uncertainty of the percentage of the network nodes seen at the end of a \gls{gathering} with an increase of the threshold for the average number of newly discovered nodes in a \gls{draw}.}\label{fig:percentage-of-network-seen-with-increasing-message-size-threshold}
\end{figure}

% Interpretation of results.
\cref{fig:percentage-of-network-seen-with-increasing-message-size-threshold} shows the average of the percentage of the network discovered with an increase of the threshold for the average number of newly discovered nodes. The shaded region around the average represents uncertainty and is bounded by one standard deviation. The experiment confirms the expected behavior, i.e., larger thresholds are not reliable in terms of percentage of network discovered, while smaller thresholds provide a more consistent result, regardless of the number of nodes known by the first contact node.

% AVG 15.0,0.9795614283467045
% MED 15.0,0.9826962403649521
% STDEV 15.0,0.009113267168597564

\begin{figure}[!t]
  \centering
  \includegraphics[width=\linewidth]{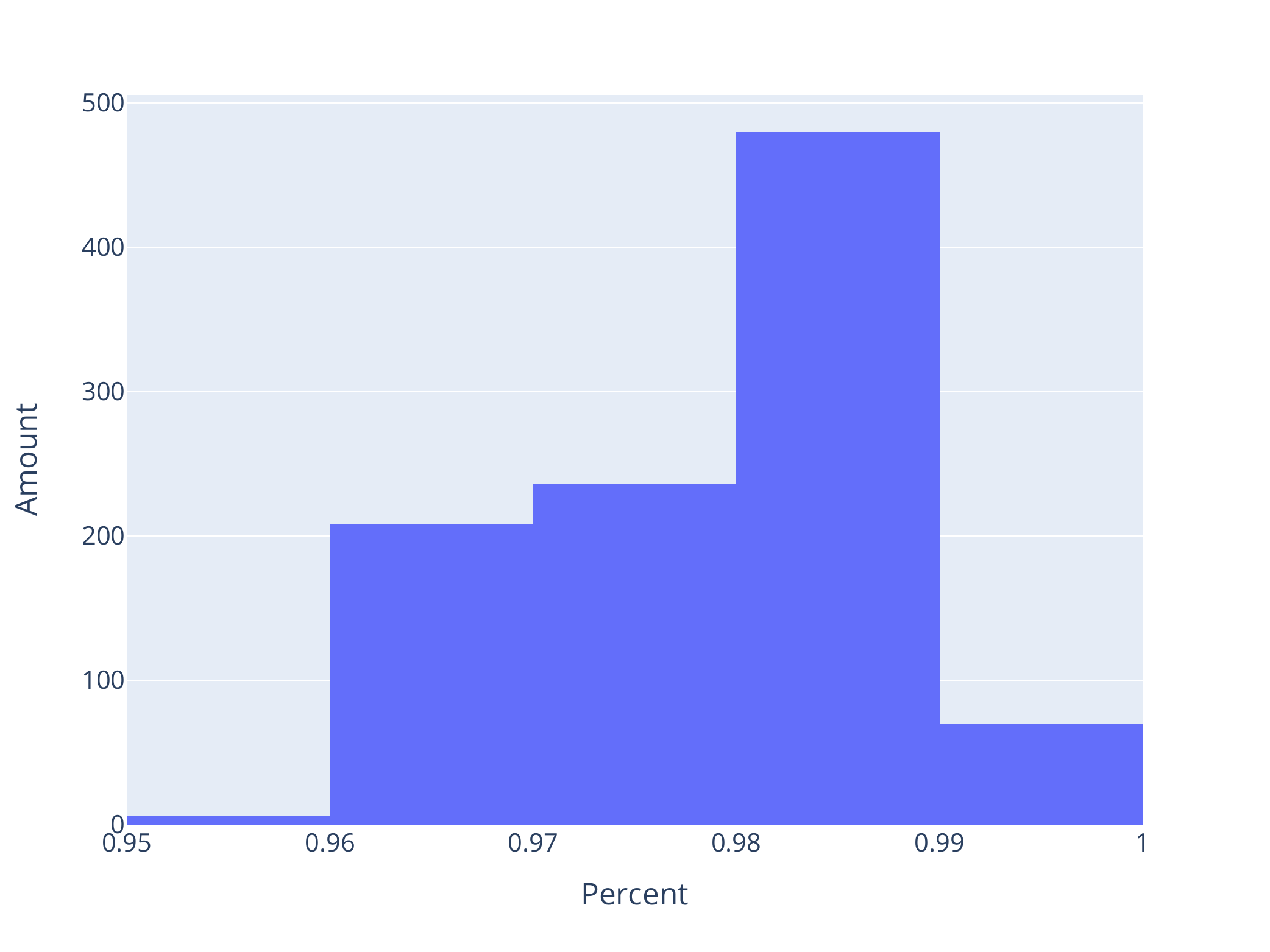}%
  \caption{Percentage of network nodes seen when the threshold on the average number of newly discovered nodes in a \gls{draw} is set to 15. }\label{fig:avg-message-threshold-15-histogram}
\end{figure}

Since~\cref{fig:percentage-of-network-seen-with-increasing-message-size-threshold} shows that when introducing a minimal threshold on the average number of newly discovered nodes in a \gls{draw} results with a high percentage of the network nodes discovered, we continue by examining a setup when the threshold is set to \(15\) (i.e., the node running the algorithm must discover on average at least \(15\) new nodes after each \gls{draw}). When such a threshold is chosen, the node discovers on average \(98.00\%\) of network nodes with a standard deviation of \(0.911\%\), as shown in~\cref{fig:avg-message-threshold-15-histogram}. This particular threshold is used in the following experiments, as we consider the result beneficial for a real world usage scenario leading to a reliable result in terms of the percentage of network nodes discovered. Note that the threshold chosen in the scope of this experiments is not applicable to any network. The method for obtaining the threshold is generic and should be repeated for a new network topology.

\subsection{Experiment 4: The upper bound on the number of generated messages}\label{exp:upper-bound-on-the-number-of-generated-messages}

% The aim of the experiment.
This experiment tests the validity of the theoretical results for communication complexity expressed by~\cref{eqn:total-number-of-exchanged-messages-before-a-decision-can-be-made} in a real network with malicious nodes. Hence the experiment investigates the number of messages generated during each simulation run.

% Execution Methodology.
We ran two simulation scenarios and for each scenario \(1000\) experiments were performed where node \gls{new-node} enters the network using a randomly-selected first contact node \gls{first-contact}. The threshold on the average number of newly discovered nodes per \gls{draw} is set to \(15\) in both scenarios\footnote{Note that this value has the same semantics as \gls{average-unique-nodes-per-draw-random-variable}, which is used in~\cref{eqn:total-number-of-exchanged-messages-before-a-decision-can-be-made}.}. In each experiment, the node running the algorithm attempts to construct and query a probabilistic progress set \gls{pps}. The construction and use of a probabilistic safe set \gls{pss} is intentionally omitted here, as it is much more difficult to construct a \gls{pps}, making \gls{pps} more suitable for this experiment, which can be viewed as a stress test. Nevertheless,~\cref{eqn:total-number-of-exchanged-messages-before-a-decision-can-be-made} can be applied regardless of the probabilistic set type (safe of progress). During the experiment we count the number of messages exchanged before the algorithm terminates, either because it managed to progress or because it halted.

In the first scenario, we set \(\gls{desired-malicious-node-tolerance}=\gls{number-of-malicious-nodes-in-the-network}=1272\). In this setting, the probabilistic set \gls{pps} can be constructed and~\cref{con:sqrt-psi-max-size} can be met quite easily.~\cref{eqn:total-number-of-exchanged-messages-before-a-decision-can-be-made} gives us an upper bound of \(728\) exchanged messages.
In the second scenario, we set \(\gls{desired-malicious-node-tolerance}=\gls{number-of-malicious-nodes-in-the-network}=1614\). Compared to the first setting, this one is more demanding: a node must collect at least \(6000\) total nodes in a \gls{gathering} in order to satisfy~\cref{con:sqrt-psi-max-size}.~\cref{eqn:total-number-of-exchanged-messages-before-a-decision-can-be-made}
gives us an upper bound of \(880\) exchanged messages.

\begin{figure}[ht!]
  \centering
  \includegraphics[width=\linewidth]{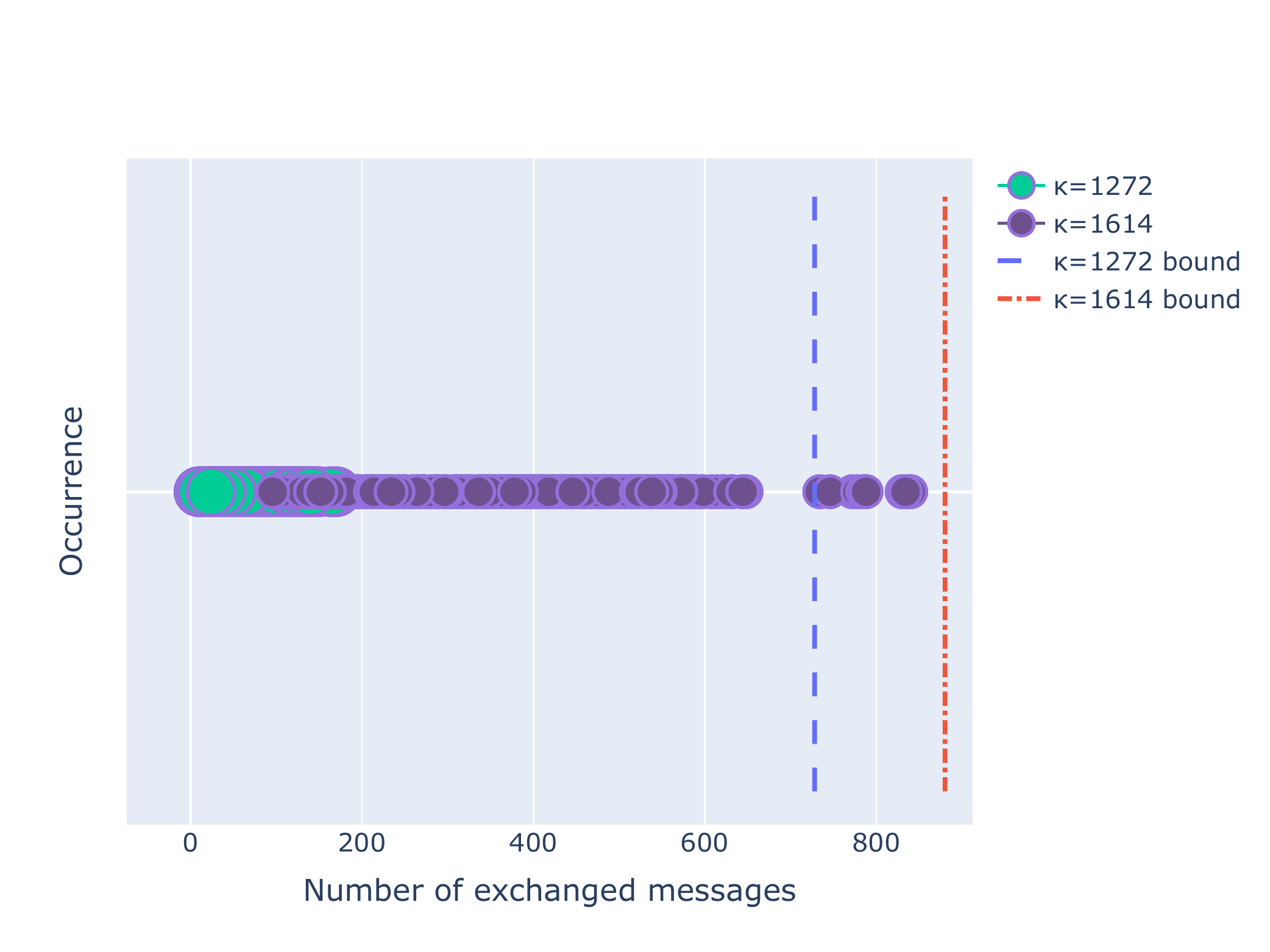}%
  \caption{Comparing the actual number of generating messages with the analytical bound on communication complexity. Markers (dots) mark one or more \glspl{gathering} that ended with the corresponding number of exchanged messages. }\label{fig:upper-bound-complexity-verification}
\end{figure}

% Interpretation of results.
The results are shown in~\cref{fig:upper-bound-complexity-verification}. In all simulation runs, the actual number of generated messages is smaller then or equal to the corresponding analytic bound, which strongly suggests that~\cref{eqn:total-number-of-exchanged-messages-before-a-decision-can-be-made} is valid.

\subsection{Experiment 5: Transaction inclusion verification for light clients}\label{sec:transaction-inclusion-verification-for-light-clients}

% The aim of the experiment.
This experiment investigates the effectiveness of the algorithm on a Bitcoin-like network containing a total of \(6356\) nodes, with a gradual increase in the total number of malicious nodes, given that the node running the algorithm does not have unlimited time and resources. Thus, this experiment examines a realistic usage scenario in which, for instance, a user has only a resource-constrained device (e.g., a mobile phone), but still wants to verify that a transaction is contained in a ledger and is interested in doing so efficiently and trustworthy.

% Execution Methodology.
The simulation setup and methodology is the same as in~\cref{exp:efficiency}, with two key differences. First,~\cref{con:sqrt-psi-max-size} must be satisfied --- if the node executing the algorithm is unable to respect the given constraint, it will not construct a probabilistic set and will consequently halt. Second, the halting condition is changed to take into account the threshold for the average number of unique nodes per \gls{draw}, which is set to \(15\), but only if \(10\) or more \glspl{draw} were previously made.

\begin{figure}[ht!]
  \centering
  \includegraphics[width=\linewidth]{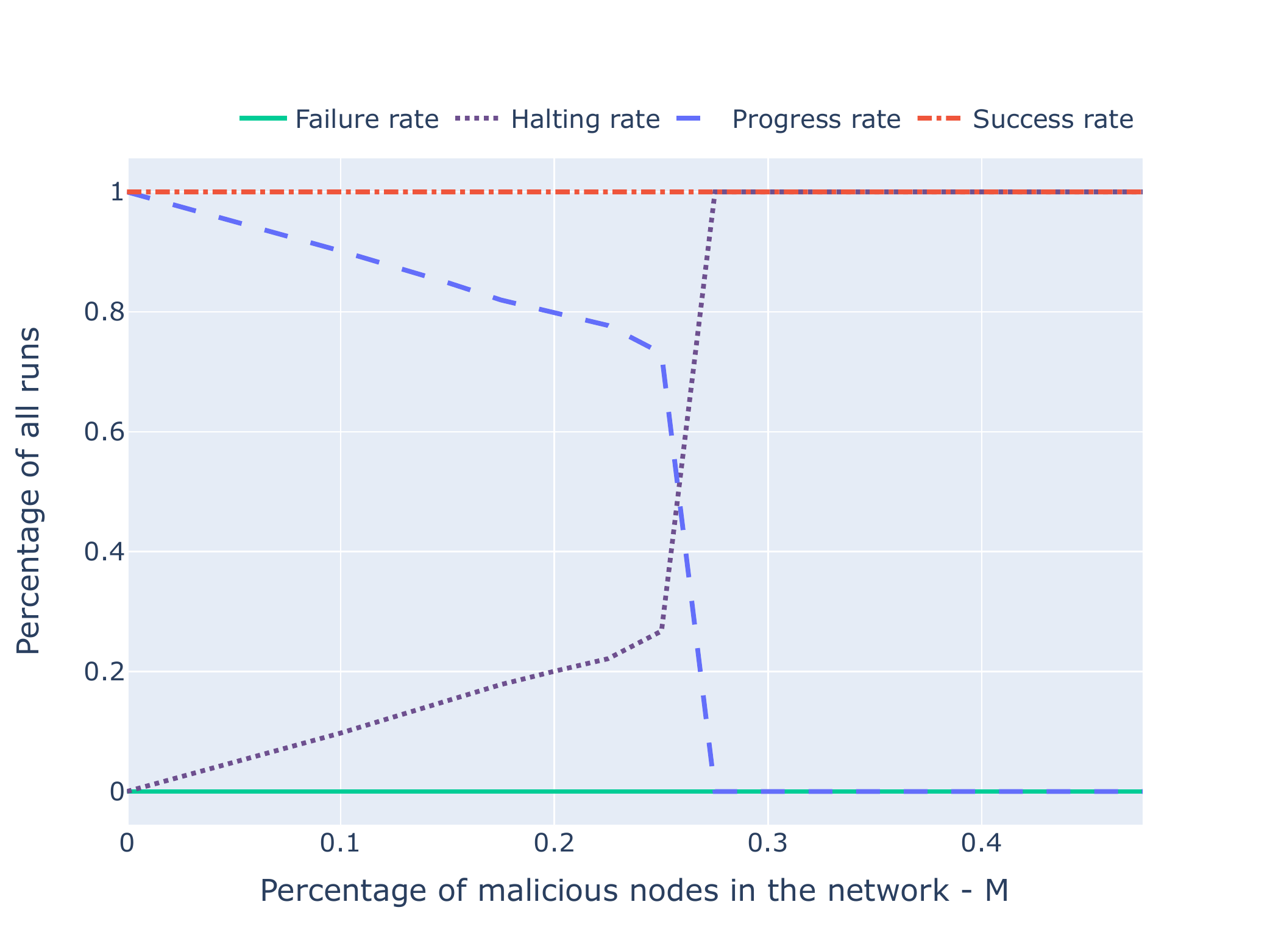}%
  \caption{Outcomes of the algorithm in a realistic setup  when~\cref{con:sqrt-psi-max-size} is satisfied and the average message size threshold is set to \(15\), while increasing the  malicious number of network nodes. }\label{fig:light-device-transaction-verification-with-the-increase-of-malicious-nodes-in-the-network}
\end{figure}

% Interpretation of results.
The results shown in~\cref{fig:light-device-transaction-verification-with-the-increase-of-malicious-nodes-in-the-network} confirm that the algorithm executes within the expected success rate with good progress while the number of reachable nodes which are malicious is below one quarter. Similarly to~\cref{fig:outcomes-with-the-increase-of-malicious-nodes-in-the-network}, the algorithm behaves as expected and identifies an honest node for transient or persistent communication in \(99.9\%\) of the trials, while halting consistently when less than a quarter of nodes in \gls{unique-nodes-found-in-gathering} are malicious. Thus, we conclude that the algorithm is adequate for resource-constrained devices when similar network conditions can be expected.

% Outro.
After verifying our work on a Bitcoin-like network, we draw three main conclusions. First, the algorithm behaves as expected, i.e., it manages to download a correct ledger, infer a state of a transaction, or halt in \gls{correctness-probability} percent of the cases. Second, since the number of detected nodes in pong messages decreases as the number of steps increases, it is useful to reduce the number of steps executed in a \gls{gathering}. This can be done by halting the \gls{gathering} when the average number of newly discovered nodes in a draw falls below a threshold. Third, the algorithm can be applied on a resource-constrained device in a realistic Bitcoin-like network when about a quarter of the network nodes are malicious. Comparing the deterministic and probabilistic variants, the probabilistic approach generates one to two orders of magnitude fewer messages when inferring the state of the ledger.

\section{Challenges and future work}\label{sec:challenges-and-future-work}

% Challenges - not all nodes available.
The most pressing challenge in a production environment concerns the responses to ping requests where network nodes are expected to respond with their peer list. Currently in Bitcoin, only publicly reachable nodes with a free slot send such replies~\cite{wang2017towards}. We can approach the problem in three different ways. First, as part of a temporary solution, if a node does not respond because a free slot is unavailable, we can filter out that node and contact another node. This will generate additional steps for the algorithm. The second solution requires a change in the node protocol so that nodes respond to ping messages regardless of the number of free slots. Such a change would not affect the underlying consensus protocol and can be presented as an opt-in feature. However, such a change makes nodes vulnerable to denial-of-service attacks. If an attack is detected, the affected node may simply decide not to respond to ping requests for a while. Third, as a persistent solution, it is likely that more publicly reachable nodes will emerge as DLT becomes more widespread and mature. As a result, more peers will be available (e.g., each BTC node can connect to 8 outgoing peers and up to 117 incoming peers~\cite{mivsic2019modeling, wang2017towards}), to partially mitigate the problem.

% Challenges - not all nodes have necessary capabilities.
%In addition, not every node currently has the necessary capabilities to run the algorithm, but these capabilities are alongside the natural technology development path and work is underway\footnote{EIP-1186: https://github.com/ethereum/EIPs/issues/1186}.

% Challenges - avoid unnecessary \glspl{gathering}.
Furthermore, one may argue that the \gls{gathering} process is resource intensive and time consuming. However, it provides trustless and decentralized capabilities which are otherwise unavailable in today's DLT networks. The caching of received peer lists and their reuse can help to mitigate the problem, while a \gls{gathering} can also be performed while the device is idle or charging. Sets can also be shared by social consensus (e.g., by a trusted person, friend, family member, etc.) and then used for future queries.

% Challenges - Incentives.
Finally, there is the question of incentive. Why would a node act as a serving node for other nodes running the Aurora protocol? The information about neighboring peers and the ledger head is already shared voluntarily by nodes, however with a limit on the maximum number of concurrent node connections. Furthermore, responses regarding the inclusion of transactions can be classified as simple requests comparable to peer list responses, so we assume that a node would be willing to share such information without additional incentives. If at any point our assumption proves to be wrong, the solution can be adapted to provide incentives for full nodes to collaborate, e.g., in the form of micro-payments for each transaction inclusion request.

% Future work.
As our future work, some of the open questions have already been mentioned, namely, the impact of \gls{correctness-probability} on the size of probabilistic sets and the use case, and the study of a time-based halting condition. We also want to study the impact of mining on the execution of the algorithm. We could compensate for new blocks generated during the execution of the algorithm by querying for blocks beyond the ledger head and then checking if there is a common ancestor. In addition, the algorithm would benefit from parallel execution of the \gls{gathering}, the effects of which can be studied in a controlled environment using a prominent distributed ledger client modified to run the Aurora algorithm.

\section{Conclusion}\label{sec:conclusion}

%  Ovdje prvo predstaviti Aurora algoritam koji ima za cilj pronaći vjeroratnosne skupove u kojima je barem 1 ili k+1 sikreni čvor. In onda navesti primjenu
The Aurora algorithm presented in this paper allows a new node entering a DLT network that contains \(|\gls{number-of-malicious-nodes-in-the-network}|\) malicious nodes to discover sets of nodes that have a high probability of containing at least \(1\) or \(1+|\gls{number-of-malicious-nodes-in-the-network}|\) honest nodes. Such sets can then be used by a new node to synchronize with the latest ledger or to determine the state of ledger transaction.

% Algorithm gist.
The algorithm can be divided into two main parts. First, the algorithm constructs a Directed Acyclic Graph over a network topology for the purpose of node discovery using response messages from contacted nodes conveying lists of their neighbors. Second, the algorithm selects a subset of the collected nodes to be tolerant to \gls{desired-malicious-node-tolerance} malicious nodes with probability \gls{correctness-probability} by relying on the hypergeometric distribution.

% Our solution to bootstrapping.
In this work, we study a scenario where bootstrap nodes are unavailable or malicious, i.e., their behavior is hostile and they conspire against a victim that enters the network. The goal of such an adversary can range from denial of service (i.e., wasting the victim's resources) to withholding the canonical truth (i.e., the longest chain) from the victim. In such an environment, we enable decentralized and trustless ledger synchronization and attempt to reduce the victim's resource consumption by providing a probabilistic algorithm that detects an honest peer with probability \gls{correctness-probability} for subsequent interaction. Specifically, the algorithm either allows ledger synchronization to continue if it determines that an honest bootstrap candidate has been found, or aborts the process if no such guarantee can be made.

% Our solution to transaction inclusion verification.
The second application of the Aurora algorithm allows a device, under the same conditions and with the same guarantees as for trustless ledger synchronization, to verify that a transaction has been included in the ledger without having to download the entire ledger or ledger header, or to trust a central authority.

% Briefly review the contents of the paper.
We compare the Aurora algorithm with existing solutions, provide the pseudocode of the algorithm, explain the prerequisites for running the algorithm, provide an analytical expression for communication complexity, and highlight two prominent applications of the algorithm. We evaluate the algorithm using Monte Carlo simulations on a Bitcoin IPV4 network slice. We used an open-source and discrete event-driven blockchain simulator to run experiments which confirmed that the experimental results are consistent with our analytical findings.

% Interpretation of results and verification.
Our experimental results show that the Aurora algorithm ensures decentralized and trustless DLT interaction of nodes running on both consumer-grade hardware and resource-constrained devices. In a realistic scenario where up to a quarter of the nodes in the network are malicious and actively attempt to subvert a new node joining the network, the algorithm is able to correctly synchronize the ledger state, infer the state of a transaction, or otherwise halt with high probability guarantees. Comparing our probabilistic algorithm with its deterministic variant, the probabilistic approach is much more efficient and generates one to two orders of magnitude fewer messages.

%TLDR
Our solution thus lowers the barrier to entry that DLT imposes on consumer-grade hardware and improves network decentralization, while allowing changes to be implemented in existing solutions without changing to the consensus protocol.

% if have a single appendix:
%\appendix[Proof of the Zonklar Equations]
% or
%\appendix  % for no appendix heading
% do not use \section anymore after \appendix, only \section*
% is possibly needed

% use appendices with more than one appendix
% then use \section to start each appendix
% you must declare a \section before using any
% \subsection or using \label (\appendices by itself
% starts a section numbered zero.)
%

\appendices

% use section* for acknowledgment
\ifCLASSOPTIONcompsoc
  % The Computer Society usually uses the plural form
  \section*{Acknowledgments}
\else
  % regular IEEE prefers the singular form
  \section*{Acknowledgment}
\fi

This work has been supported in part by Croatian Science Foundation under the project IP-2019-04-1986 (IoT4us: Human-centric smart services in interoperable and decentralized IoT environments).

% Can use something like this to put references on a page
% by themselves when using endfloat and the captionsoff option.
\ifCLASSOPTIONcaptionsoff{}
\newpage
\fi

% trigger a \newpage just before the given reference
% number - used to balance the columns on the last page
% adjust value as needed - may need to be readjusted if
% the document is modified later
%\IEEEtriggeratref{8}
% The "triggered" command can be changed if desired:
%\IEEEtriggercmd{\enlargethispage{-5in}}

% references section

% can use a bibliography generated by BibTeX as a .bbl file
% BibTeX documentation can be easily obtained at:
% http://mirror.ctan.org/biblio/bibtex/contrib/doc/
% The IEEEtran BibTeX style support page is at:
% http://www.michaelshell.org/tex/ieeetran/bibtex/
%\bibliographystyle{IEEEtran}
% \bibliographystyle{IEEEtran}
% argument is your BibTeX string definitions and bibliography database(s)
%\bibliography{IEEEabrv,../bib/paper}
% \bibliography{library}
%
% <OR> manually copy in the resultant .bbl file
% set second argument of \begin to the number of references
% (used to reserve space for the reference number labels box)
% Generated by IEEEtran.bst, version: 1.14 (2015/08/26)

% biography section
% 
% If you have an EPS/PDF photo (graphicx package needed) extra braces are
% needed around the contents of the optional argument to biography to prevent
% the LaTeX parser from getting confused when it sees the complicated
% \includegraphics command within an optional argument. (You could create
% your own custom macro containing the \includegraphics command to make things
% simpler here.)
%\begin{IEEEbiography}[{\includegraphics[width=1in,height=1.25in,clip,keepaspectratio]{mshell}}]{Michael Shell}
% or if you just want to reserve a space for a photo:

\begin{IEEEbiography}
  [{\includegraphics[width=1in,height=1.25in,clip,keepaspectratio]{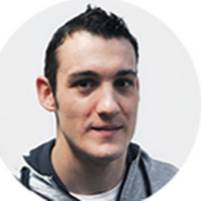}}]
  {Federico Matteo Benčić} received the M.Sc.\ degree in information and communication technology from the University of Zagreb, Faculty of Electrical Engineering and Computing in 2017, where he is currently pursuing the Ph.D. degree with the Department of Telecommunications. He is currently an Assistant with the Department of Telecommunications, Faculty of Electrical Engineering and Computing, University of Zagreb. His interests include applications of DLT in the IoT area.
\end{IEEEbiography}

\begin{IEEEbiography}
  [{\includegraphics[width=1in,height=1.25in,clip,keepaspectratio]{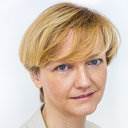}}]
  {Ivana Podnar Žarko} is Full Professor at the University of Zagreb, Faculty of Electrical Engineering and Computing, Croatia (UNIZG-FER) where she teaches distributed information systems. She received her B.Sc., M.Sc.\ and Ph.D. degrees in electrical engineering from UNIZG-FER, in 1996, 1999 and 2004, respectively. She is affiliated with the Department of Telecommunications at UNIZG-FER from 1997. She was a guest researcher and research associate at the Technical University of Vienna, Austria, and a postdoctoral researcher at the Swiss Federal Institute of Technology in Lausanne (EPFL), Switzerland. She was promoted to Full Professor in December 2017. She has participated in a number of research projects funded by national sources and EU funds, and is currently leading the UNIZG-FER Internet of Things Laboratory. Ivana Podnar Žarko is the Technical Manager of the H2020 project symbIoTe: Symbiosis of smart objects across IoT environments, and is currently participating in the Centre of Research Excellence for Data Science and Advanced Cooperative Systems, which is the first national center of excellence in the field of technical sciences in Croatia. She has co-authored more than 60 scientific journal and conference papers in the area of large-scale distributed systems, IoT, and Big data processing, and has as a program committee member for a number of international conferences and workshops. Prof.\ Ivana Podnar Žarko is a member of IEEE and was the Chapter Chair of IEEE Communications Society, Croatia Chapter (2011--- 2014). She has received the award for engineering excellence from the IEEE Croatia Section in 2013.
\end{IEEEbiography}

% insert where needed to balance the two columns on the last page with
% biographies

% You can push biographies down or up by placing
% a \vfill before or after them. The appropriate
% use of \vfill depends on what kind of text is
% on the last page and whether or not the columns
% are being equalized.

%\vfill

% Can be used to pull up biographies so that the bottom of the last one
% is flush with the other column.
%\enlargethispage{-5in}
\clearpage
% \glsaddall{}
\printglossary{}
% that's all folks
\end{document}